# 8

# Saturn's Ionosphere

L. MOORE[1], M. GALAND[2], A.J. KLIORE[3], A.F. NAGY[4], AND J. O'DONOGHUE[5]

[1] Center for Space Physics, Boston University, Boston, MA, USA.
[2] Space and Atmospheric Physics Group, Department of Physics, Imperial College London, London, UK.
[3] Jet Propulsion Laboratory, California Institute of Technology, Pasadena, CA, USA.
[4] Department of Climate and Space Sciences and Engineering, University of Michigan, Ann Arbor, MI, USA.
[5] NASA Goddard Space Flight Center, Greenbelt, MD, USA.

**Abstract**

This chapter summarizes our current understanding of the ionosphere of Saturn. We give an overview of Saturn ionospheric science from the Voyager era to the present, with a focus on the wealth of new data and discoveries enabled by Cassini, including a massive increase in the number of electron density altitude profiles. We discuss recent ground-based detections of the effect of "ring rain" on Saturn's ionosphere, and present possible model interpretations of the observations. Finally, we outline current model-data discrepancies and indicate how future observations can help in advancing our understanding of the various controlling physical and chemical processes.

## 8.1 Introduction

Saturn's upper atmosphere is typically defined to be the region above the homopause, which marks the transition between a well-mixed atmospheric region dominated by eddy diffusion below (the lower atmosphere; Chapter 14) and a region dominated by molecular diffusion above. It can further be broken down into two coincident regions: the neutral thermosphere (Chapter 9), and the charged ionosphere. The upper atmosphere forms the transition region between a dense neutral atmosphere below and a tenuous, charged magnetosphere above (Chapter 6); consequently it also mediates the exchange of particles, momentum, and energy between these two regions. External forcing on the upper atmosphere, such as by solar extreme ultraviolet (EUV) photons or energetic particles, determines the degree of ionization within the ionosphere. As magnetic fields strongly influence charged particle motions, the ionosphere tends to be more strongly ionized where magnetic field configurations favor precipitation of energetic particles into the atmosphere; namely, the auroral regions near the magnetic poles (Chapter 7).

In this chapter we summarize our current understanding of the non-auroral ionosphere of Saturn, drawing from ground-based and space-based measurements (Section 8.2) and from modeling studies for interpreting the observations and physical processes driving Saturn's ionosphere (Section 8.3). As giant planet ionospheres are all qualitatively similar we try to avoid reproducing material unnecessarily from previous giant planet review articles. Therefore, those reviews are still highly relevant (e.g., *Atreya et al.*, 1984; *Waite et al.*, 1997; *Nagy and Cravens*, 2002; *Majeed et al.*, 2004; *Yelle and Miller*, 2004; *Nagy et al.*, 2009; *Schunk and Nagy*, 2009). While we will touch upon closely related topics as appropriate, more thorough discussions of Saturn's aurorae and thermosphere can be found in chapters 7 and 9, respectively.

## 8.2 Observations

There are several options for remotely observing Saturn's ionosphere, and each technique has not only particular advantages but also significant, unique limitations. To date, the vast majority of observational information regarding Saturn's ionosphere has come from spacecraft radio occultations, which yield vertical electron density structure, $N_e(h)$. Due to Sun-Saturn-Earth geometry, however, a spacecraft-to-Earth radio occultation can only be performed near the terminator and therefore samples only the dawn or dusk ionosphere of Saturn (see Section 8.2.1). An additional method of tracking the peak ionospheric electron density, $N_{MAX}$, involves the detection of broadband radio emission (dubbed Saturn Electrostatic Discharge, SED) originating from powerful lightning storms in Saturn's lower atmosphere. As the storm rotates, a nearby spacecraft can use SED emission to derive the diurnal variation of $N_{MAX}$ near the storm location. Unfortunately this technique is reliant upon storm activity and latitude, and it does not contain any altitude information (Section 8.2.2). Finally, infrared measurements of rotational-vibrational emissions from $H_3^+$ near 3-4 μm yield temperature and density information for this major ion in Saturn's ionosphere. Aside from one so-far-unique observation, however, $H_3^+$ emission has proven to be too weak to be detected outside of Saturn's auroral regions (Section 8.2.3).

### 8.2.1 Radio Occultations

The technique of radio occultation, whereby Saturn's atmosphere occults the transmission of a radio signal from a spacecraft to Earth (e.g., *Lindal*, 1992; *Kliore et al.*, 2004), provides the only available remote diagnostic of electron density altitude profiles, $N_e(h)$, a basic ionospheric property. There are, at most, two opportunities for deriving atmospheric properties during a spacecraft flyby: one during the occultation ingress (or entry, often designated with "N") and one during the occultation egress (or exit, often designated with "X"). The geometry for essentially all of the radio occultations performed at Saturn to date is such that the ingress occultations sample the dusk ionosphere while the egress occultations sample the dawn ionosphere. Radio occultation latitude, typically given in planetographic coordinates, refers to the latitude at the lowest altitude of the occultation. As the spacecraft almost never follows a trajectory in which the occultation ray path is continually above a single latitude, however, the altitude profile derived from a radio occultation actually samples a range of latitudes. Quoted radio occultation latitudes therefore usually refer to either an approximate latitude or to the latitude at the base of the occultation near the electron density peak.

The main approach for analyzing spacecraft radio occultations is based on geometrical optics (e.g., *Fjeldbo et al.*, 1971). In this approach, the deviation of a ray path in response to refractive index gradients in ionospheric plasma is tracked, leading ultimately to a Doppler shift in the frequency of the signal received at Earth. The time series of differences between the transmitted and received frequencies, called the frequency residuals, can then be used to derive a vertical profile of refractive index and electron number density (e.g., *Withers et al.*, 2014). This direct approach is particularly susceptible to multipath propagation effects, wherein narrow ionospheric layers can lead to multiple, distinct signals arriving simultaneously at the receiving antenna, each with a different Doppler-shifted frequency. This effect is stronger the farther a spacecraft is away from the occulting planet. A better approach is to first use scalar diffraction theory, which transforms the data using Fourier analysis in order to mimic an occultation by a nearby spacecraft, thereby removing complications due to multipath propagation and diffraction effects (e.g., *Karayel and Hinson*, 1997; *Hinson et al.*, 1998). Once this is accomplished the geometric approach can then be used. Saturn radio occultations have so far only been analyzed by the geometrical optics technique and therefore are highly uncertain in the presence of narrow ionospheric layers, which appear to be particularly common at lower altitudes. In addition to instrumental effects, other important sources of uncertainty include: the required assumption of spherical symmetry within the ionosphere, the degree of intervening plasma between the transmitting and receiving antenna, and inaccurate positions or velocities. The propagation of uncertainties through all of the processing steps is non-trivial (e.g., *Lipa and Tyler*, 1979), but typical estimates are on the order of a few hundred electrons per cubic centimeter for Cassini radio occultations (*Nagy et al.*, 2006; *Kliore et al.*, 2011).

Flybys of the Saturn system by Pioneer 11 (1 September 1979), Voyager 1 (12 November 1980) and Voyager 2 (26 August 1981) yielded our first observational insights into Saturn's ionospheric electron densities (*Kliore et al.*, 1980; *Lindal et al.*, 1985). While there was some variation in the electron density profiles obtained by the Pioneer 11 and Voyager spacecraft, the altitudes of the electron density peaks ($h_{MAX}$) ranged from ~1000-2800 km and the maximum electron density ($N_{MAX}$) of 5 out of the 6 profiles was of order $10^4$ cm$^{-3}$. The Voyager 2 ingress profile was an outlier from this trend, with narrow low-altitude layers of peak electron density near $7 \times 10^4$ cm$^{-3}$. Such layers appear to be relatively common in the giant planet ionospheres, as they have also been found at Jupiter (e.g., *Fjeldbo et al.*, 1975; *Yelle and Miller*, 2004), Uranus (*Lindal et al.*, 1987), and Neptune (*Lindal*, 1992). While there is a lack of consensus regarding the origin of these layers, a likely explanation is that vertical wind shears – such as those that might result from atmospheric gravity waves – can lead to localized electron density enhancements. This effect has been demonstrated through modeling of Saturn's upper atmosphere by *Moses and Bass* (2000) and *Barrow and Matcheva* (2013), and also of Jupiter (*Matcheva and Strobel*, 2001) and Neptune (*Lyons*, 1995).

The arrival of Cassini at Saturn in 2004 has significantly increased the number of radio occultations, allowing for a more thorough examination of possible ionospheric trends. A total of 59 Cassini radio occultations have already been obtained and analyzed, with one final occultation currently planned (*Kliore et al.*, 2014). The first dozen Cassini radio occultations were obtained between May and September 2005 and sampled Saturn's near-equatorial region between 10°N and 10°S planetographic latitude. These profiles, shown in **Figure 8.1**, revealed a clear dawn-dusk asymmetry. While there is still a high degree of variability among the 12 profiles, on average the peak densities are lower and the peak altitudes are higher at dawn than at dusk (*Nagy et al.*, 2006). This trend is also present in the complete Cassini radio occultation dataset, which includes a number of additional low-latitude profiles (*Kliore et al.*, 2014). Such a behavior is consistent with the expectation that chemical losses during the Saturn night would lead to a depletion of the low-altitude ionospheric electron density peak. The averages of the first dozen low-latitude Cassini profiles (5 dusk and 7 dawn) are shown in **Figure 8.2**.

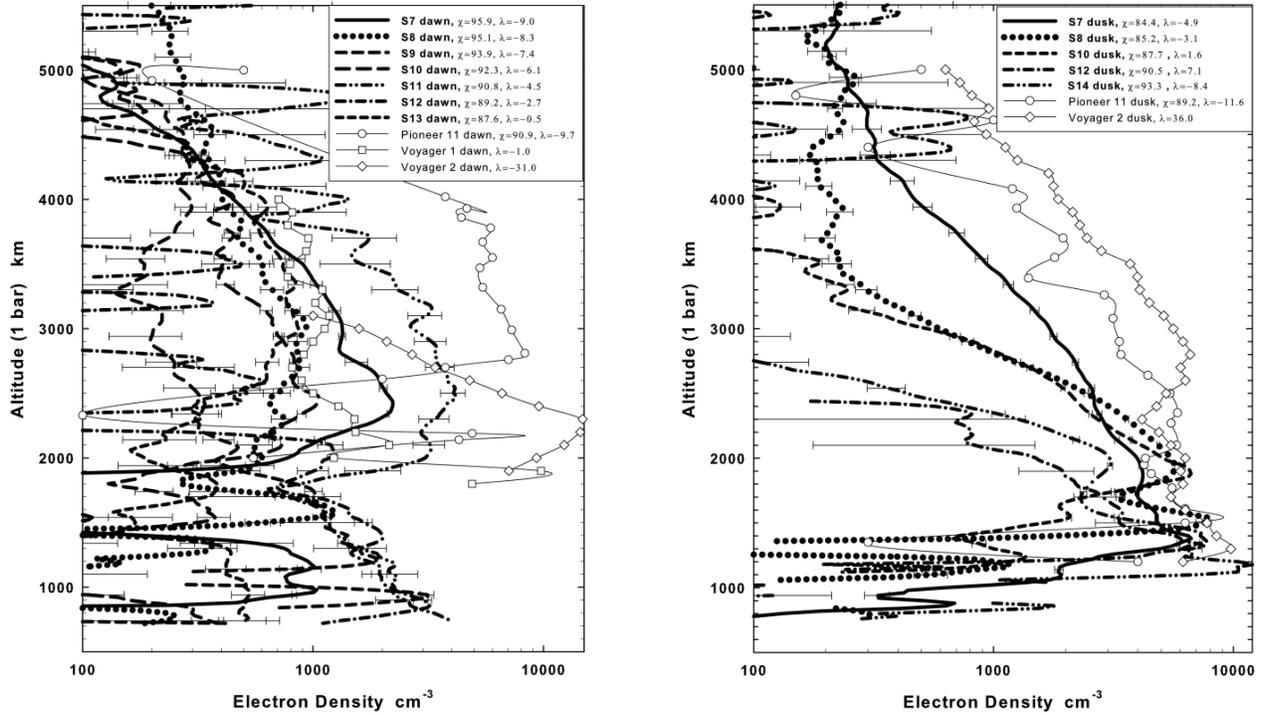

**Figure 8.1:** Saturn ionospheric electron density altitude profiles retrieved from radio occultations by the Cassini spacecraft at **(left)** dawn and **(right)** dusk. Also shown for comparison are Pioneer 11 and Voyager radio occultation profiles. Error bars represent the uncertainties introduced by baseline frequency fluctuations and the effects of averaging data from multiple Deep Space Network stations. All Cassini profiles are from ±10° latitude. From *Nagy et al.* (2006).

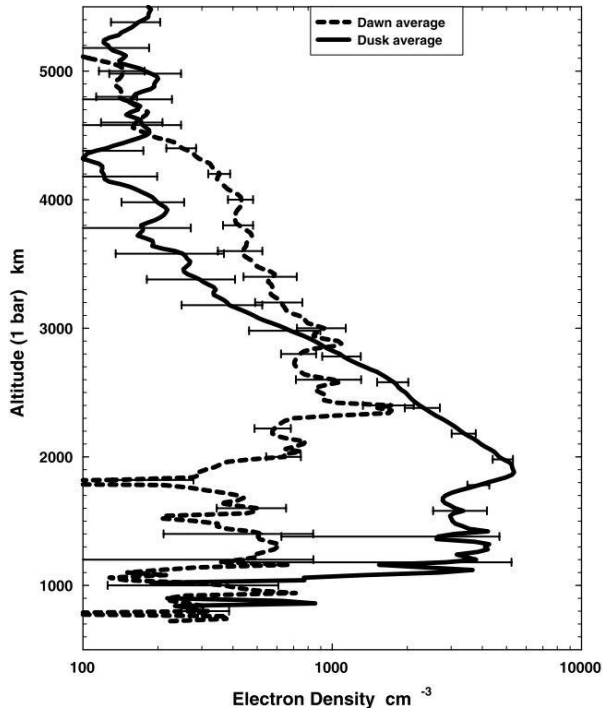

**Figure 8.2:** Weighted averages of the dawn and dusk electron density profiles from Figure 8.1. From *Nagy et al.* (2006).

Subsequent Cassini radio occultations sampled a wider range of latitudes, revealing a surprising trend in $N_{MAX}$. Nineteen low-, mid- and high-latitude radio occultations obtained between September 2006 and July 2008 found that peak electron densities were smallest at Saturn's equator, and increased with latitude (*Kliore et al.*, 2009). The average sub-solar latitude during this occultation period was -8.5°, meaning that the sun primarily illuminated Saturn's low latitudes. Solar EUV photons are expected to be the primary source of ionization in Saturn's non-auroral ionosphere, and yet Cassini radio occultations revealed the minimum electron densities to be in regions of maximum insolation.

This latitudinal electron density trend was reconfirmed and bolstered with 28 additional occultations obtained between 2008 and 2013 (*Kliore et al.*, 2014). While there are a few high altitude occultations, unfortunately none of them appear to have sampled active auroral regions (*Moore et al.*, 2010).

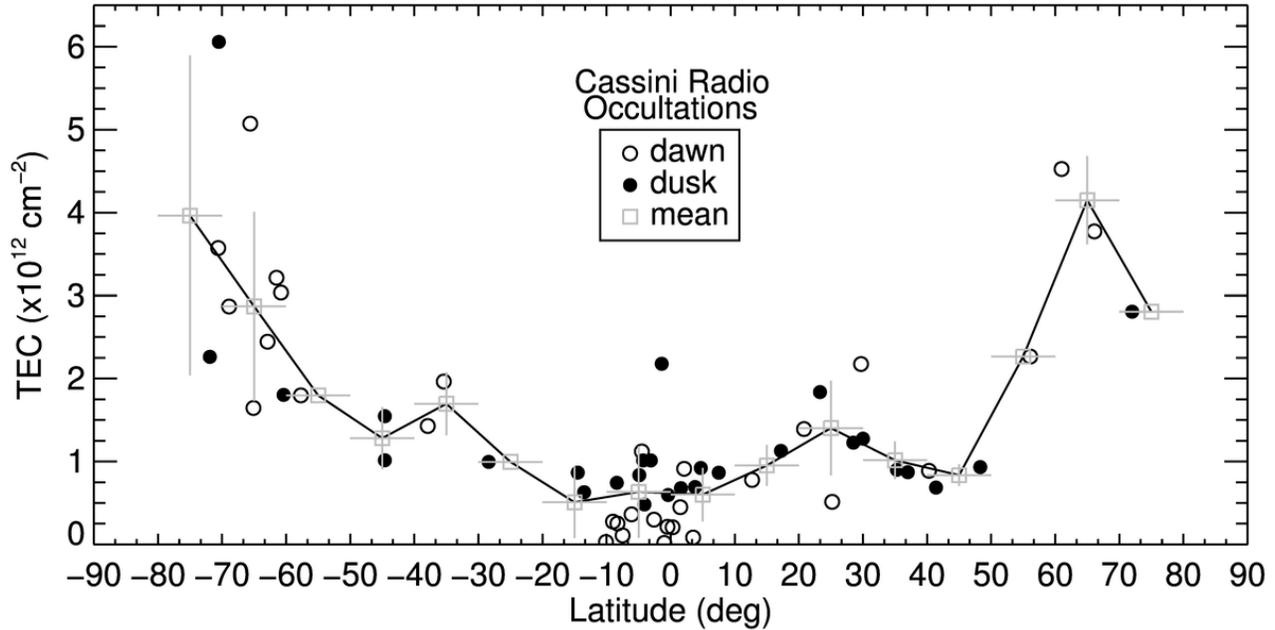

**Figure 8.3:** Total Electron Content (TEC) from all 59 published Cassini radio occultations. Dawn (exit) profiles are shown as open circles, dusk (entry) profiles are shown as filled circles. Gray squares represent the local mean over 10° latitude bins. Vertical gray bars indicate the standard deviation of the mean, while horizontal gray bars demarcate the latitude range over which the mean is calculated. Based on Table 1 of *Kliore et al.* (2014).

**Figure 8.3** shows the total electron content (TEC), or column integrated electron density, for all 59 of the published Cassini radio occultations, plotted versus planetographic latitude. The trend in TEC is quite similar to that in $N_{MAX}$, though with considerably less scatter (*Kliore et al.*, 2014). In both cases, there is a clear minimum in electron density at Saturn's equator, and an increase in electron density with latitude. There also appears to be a local minimum in electron density around 45°N latitude, near the region of Saturn's atmosphere that is magnetically linked to the inner B ring, long predicted to be the site of an enhanced influx of water from the rings to the atmosphere (e.g., *Connerney and Waite*, 1984; *Connerney*, 1986). The introduction of oxygen, whether in the form of neutral or charged water or other oxygen-bearing molecules or charged sub-micrometer grains, acts to reduce the local electron density, as it converts the long-lived atomic ion $H^+$ into a short-lived molecular ion. While it is tempting to associate the localized minimum in electron density near 45°N with an enhanced water influx, ring-derived water influxes are expected to be stronger in the southern hemisphere due to Saturn's effectively offset magnetic dipole (*Burton et al.*, 2010), independent of Saturn season (e.g., *Northrop and Connerney*, 1987; *Tseng et al.*, 2010). No similar localized minimum is obvious in the southern hemisphere radio occultations. There is a slight hint of a minimum near 45°S, though the sampling statistics are poor in that region.

Cassini radio occultation measurements have demonstrated that Saturn's ionosphere is highly variable, with electron density altitude profiles obtained at similar latitudes from similar times often differing significantly from each other, and with frequent narrow low-altitude layers of electron density. Nevertheless, the wealth of data – at least compared with other giant planet ionospheres – has also allowed for identification of a number of trends. On average, peak electron densities are smaller and peak altitudes are higher at dawn than at dusk, consistent with recombination of major ions during the Saturn night. On average, the smallest electron densities are found near Saturn's equator, and electron densities increase with latitude, contrary to what would be expected from an ionosphere driven purely by solar photoionization with constant photochemical loss sources. Finally, though the statistics are poor, radio occultation observations also give some indication that there may be localized minima in electron density near 45°N planetographic latitude, which could be consistent with an influx of charged water grains from Saturn's rings to its atmosphere along magnetic field lines.

### 8.2.2   Saturn Electrostatic Discharges (SEDs)

During Voyager encounters with Saturn, the Planetary Radio Astronomy (PRA) instrument detected mysterious, broadband, short-lived, impulsive radio emission (*Warwick et al.*, 1981, 1982). These radio bursts, termed Saturn Electrostatic Discharges (SEDs), were organized in episodes lasting several hours and separated from each other by roughly 10h 10m. While there was initially some uncertainty regarding the origin of SEDs, *Burns et al.* (1983) suggested they were radio manifestations of atmospheric lightning storms, and *Kaiser et al.* (1984) demonstrated that an extended source region in the equatorial atmosphere was consistent with the observed SED recurrence pattern. Cassini's Radio and Plasma Wave Science (RPWS) instrument began detecting SEDs prior to its orbital insertion on 1 July 2004, and has since observed nine distinct storm periods, separated by SED-quiet periods of a few days to 21 months (*Fischer et al.*, 2011b). Shortly after Cassini's arrival at Saturn, the Imaging Science Subsystem instrument detected a storm system at 35°S planetocentric latitude that correlated with the observed SED recurrence pattern (*Porco et al.*, 2005). *Dyudina et al.* (2007) extended this finding by presenting three further storm systems where SED observations were correlated with the rising and setting of a visible storm on the Saturn radio horizon. Finally, lightning flashes were imaged directly by Cassini in 2009, providing a convincing demonstration that SEDs were indeed radio signatures of atmospheric storms in Saturn's lower atmosphere (*Dyudina et al.*, 2010).

SEDs have a large frequency bandwidth, but appear as narrow banded streaks in both Voyager PRA and Cassini RPWS dynamic spectra due to the short duration of the radio burst and the frequency sampling nature of the receivers. The number of SEDs detected in an individual storm varies significantly, from hundreds to tens of thousands (*Fischer et al.*, 2008), with typical burst rates of a few hundred per hour (*Zarka and Pedersen*, 1983; *Fischer et al.*, 2006). SED *storms* are periods of nearly continuous SED activity, modulated by *episodes* of varying SED activity. The recurrence period of the episodes within a storm represents the time between peaks of SED activity; for a single longitudinally confined storm system, therefore, this period is closely related to the local rotation rate of the atmosphere.

Recurrence periods for Voyager 1 and 2 SED episodes were ~10h 10m and ~10h 00m, respectively (*Evans et al.*, 1981; *Warwick et al.*, 1982), and were consequently thought to originate from equatorial storm systems (*Burns et al.*, 1983), though none were observed directly. In contrast, the majority of recurrence periods for Cassini era SED storms are near 10h 40m (*Fischer et al.*, 2008), implying a mid-latitude origin, as confirmed by the 35°S planetocentric latitude storm clouds and visible lightning flashes imaged by Cassini. Approximately 16 months after Saturn passed through its equinox (August 2009) towards southern winter, a giant convective storm developed at 35°N planetocentric latitude, accompanied by unprecedented levels of SED activity, with flash rates an order of magnitude higher than previously observed storms (*Fischer et al.*, 2011a; this storm is described in detail in Chapter 13). While the tendency for Saturn lightning storms to preferentially form near ±35° planetocentric latitude remains unexplained, it is important to keep in mind that SEDs appear to primarily probe either Saturn's mid-latitude (Cassini era) or equatorial (Voyager era) ionosphere.

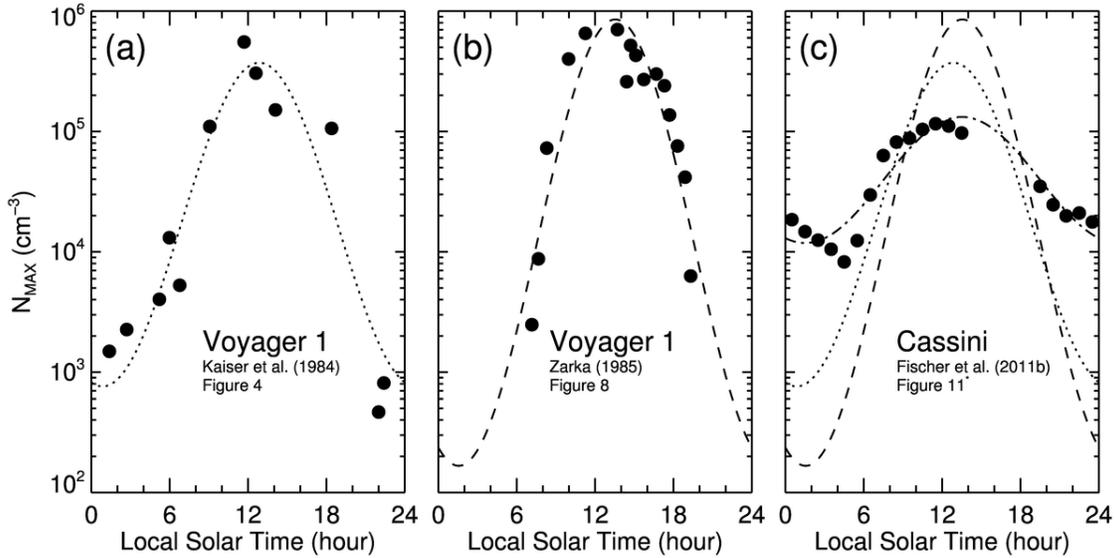

**Figure 8.4:** Diurnal variation in $N_{MAX}$ derived from Voyager 1 and Cassini SED observations (filled circles), along with a least-squares fit to an equation of the form *log $N_e$ = A – B cos(SLT – $\phi$)*, where SLT is solar local time in hours and $\phi$ is the phase shift (the dotted, dashed, and dot-dash curves). **(a)** Voyager 1, taken from Figure 4 of *Kaiser et al.* (1984); **(b)** Voyager 1, taken from Figure 8 of *Zarka* (1985); and **(c)** Cassini, based on Figure 11 of *Fischer et al.* (2011). The Voyager 1 fits (i.e., dotted and dashed curves) are repeated in (c) in order to more easily compare them with the diurnal variation derived from Cassini era SEDs (dot-dash curve). Adapted from Figure 1 of *Moore et al.* (2012).

SEDs originating from lightning storms deep within Saturn's atmosphere must ultimately transit the ionosphere in order to be detected by a spacecraft. Therefore, the low frequency cutoff of each SED episode provides information about the intervening plasma densities, as only frequencies larger than the peak electron plasma frequency will pass through Saturn's ionosphere. *Kaiser et al.* (1984) combined the observed low frequency cutoffs with a ray tracing analysis in order to derive the peak electron density along the propagation path between Voyager 1 and the SED storm. Using similar assumptions, *Zarka* (1985) also derived peak electron densities from Voyager 1 SED measurements. (Voyager 2 data showed a decline in number and intensity of SEDs with no clear episodic behavior, meaning it could not be used for a similar analysis.) Both *Kaiser et al.* (1984) and *Zarka* (1985) found that $N_{MAX}$ varied by more than two orders of magnitude throughout the Saturn day, with midnight electron densities below $10^3$ cm$^{-3}$ and noon densities greater than $10^5$ cm$^{-3}$. The SED-derived dawn and dusk $N_{MAX}$ values were of order $10^4$ cm$^{-3}$, in rough agreement with radio occultation results. These diurnal $N_{MAX}$ trends, along with a Cassini era SED-derived trend, are presented in **Figure 8.4**.

Whereas Voyager 1 $N_{MAX}$ trends were derived from three SED episodes, Cassini SED observations allowed for investigation of diurnal trends over different storm periods, different episodes within storm periods, and different Cassini-Saturn distances. The diurnal $N_{MAX}$ trend shown in **Figure 8.4c** was based on 48 SED episodes between 2004 and 2009 when Cassini was within 14 $R_S$ of Saturn (Figure 11 of *Fischer et al.*, 2011b). This profile is slightly different from the Cassini era trend that includes all of the SED observations through 2009 (Figure 9 of *Fischer et al.*, 2011b), most likely due to the fact that attenuation of radio waves by Saturn's ionosphere is frequency-dependent, and Cassini is better able to detect weaker bursts at lower frequencies when it is nearer to Saturn. In contrast to the two-order-of-magnitude diurnal variation in $N_{MAX}$ derived from Voyager SEDs, the Cassini SED-derived $N_{MAX}$ varies by only one order of magnitude, between ~$10^4$ cm$^{-3}$ and ~$10^5$ cm$^{-3}$, with a local minimum just before sunrise. Such a local minimum in electron density is consistent with the nighttime chemical loss due to recombination implied by radio occultations; a minimum at midnight is much more difficult to understand theoretically (e.g., *Majeed and McConnell*, 1996).

While peak electron densities derived from SEDs are highly complementary to the dawn/dusk electron density profiles retrieved from radio occultation measurements, there are also some unique limitations to bear in mind. First, the frequency-dependent attenuation of radio waves by Saturn's ionosphere highlights an ambiguity regarding the observed cutoff frequency. For example, *Fischer et al.* (2011b) find a correlation between Cassini-Saturn distance and cutoff frequency: on average Cassini observes lower cutoff frequencies (and therefore derives lower $N_{MAX}$ values) when it is closer to Saturn. There are relatively few SED episodes with Cassini closer than 5 $R_S$, however, so it is not clear whether this trend continues radially inwards towards Saturn. If so, it may imply that the current derived $N_{MAX}$ values should be reduced in magnitude. It is not immediately obvious how such a correlation would affect the Voyager era results, though it is worth noting that the anomalously low frequency cutoffs occurred during Voyager's closest approach (*Kaiser et al.*, 1984). These anomalous low frequency cutoffs also occurred near ~300-600 kHz, where Saturn kilometric radiation, or SKR, typically dominates the frequency spectrum, a fact that can complicate SED analysis (*Fischer et al.*, 2011b).

The second main limitation of SED observations is the uncertainty regarding the portion of the ionosphere sampled by the transiting radio waves. Over-horizon SEDs have been observed regularly by Cassini – that is, SED detections prior to the rising of their originating storm above the horizon as seen by the spacecraft (*Fischer et al.*, 2008). These SEDs are likely a result of ducting (*Zarka et al.*, 2006), wherein propagating radio waves are refracted by the ionosphere, and their detection emphasizes that one cannot rely on the assumption that SEDs traverse a straight line from their origin to the observer. Consequently it is possible that SEDs sample portions of the ionosphere different from where the radio signals originate. For example, shadowing by Saturn's rings leads to patterns of depleted electron density that depend on season, and may help explain the anomalously low frequency cutoffs observed by Voyager (*Burns et al.*, 1983; *Mendillo et al.*, 2005). Nevertheless, despite the above limitations, SEDs provide valuable insight to Saturn's ionosphere as well as an additional observational constraint that remains to be explained: the strong diurnal variation of $N_{MAX}$ (of 1-2 orders of magnitude). While weak SED-like radio spikes were detected by Voyager 2 during its encounters with Uranus (*Zarka and Pedersen*, 1986) and Neptune (*Kaiser et al.*, 1991), no high-frequency radio emission from lightning was detected at Jupiter by any visiting spacecraft, despite whistler and optical lightning detections (*Zarka et al.*, 2008), possibly due to ionospheric attenuation (*Zarka*, 1985b) or to slow lightning discharge (*Farrell et al.*, 1999).

### 8.2.3   Observations of ionospheric $H_3^+$

Since its initial discovery in the Jovian atmosphere (*Drossart et al.*, 1989), $H_3^+$ has been an effective probe of the auroral ionospheres of Jupiter (e.g., *Lystrup et al.*, 2008; *Stallard et al.*, 2012b, and references therein), Saturn (e.g., *Stallard et al.*, 2008, 2012a; *O'Donoghue et al.*, 2014, and references therein; see also Chapter 7), and Uranus (*Melin et al.*, 2013). There is an abundance of strong $H_3^+$ rotational-vibrational emission lines available in the near-IR, as described in Chapter 7. Ionization of molecular hydrogen, the dominant constituent of Saturn's upper atmosphere, leads directly to the production of $H_3^+$ via the rapid ion-molecule reaction ($H_2^+ + H_2 \rightarrow H_3^+ + H$). Therefore $H_3^+$, along with $H^+$, is expected to be a major ion in giant planet ionospheres. As $H_3^+$ is in quasi-local thermodynamic equilibrium with the neutral atmosphere in the collisional region of the ionosphere, it also serves as a valuable probe of upper atmospheric temperatures (e.g., *Miller et al.*, 2000).

At Jupiter, where ionospheric densities and upper atmospheric temperatures are relatively large, $H_3^+$ can be observed at all latitudes. Low- and mid-latitude $H_3^+$ column densities at Jupiter are approximately $\sim$3-5x$10^{11}$ cm$^{-2}$, whereas auroral column densities can be more than $10^{12}$ cm$^{-2}$ (*Lam et al.*, 1997; *Miller et al.*, 1997). Auroral column densities at Saturn have been measured to be of the same order, though variable, with reported values between $\sim$1-7x$10^{12}$ cm$^{-2}$ (*Melin et al.*, 2007). The strong dependence of $H_3^+$ emission on temperature, however, has inhibited searches for non-auroral $H_3^+$ at Saturn, due to the low equatorial temperatures described in Chapter 9. Prior to 2011 the lowest latitude of detected $H_3^+$ at Saturn was $\sim$57° (planetographic south), a weak secondary auroral oval thought to be associated with the breakdown in corotation within the magnetosphere (*Stallard et al.*, 2008, 2010). Observations made in April 2011 using the Near-InfraRed Spectrograph (NIRSPEC) instrument on the Keck telescope detected the first low- and mid-latitude $H_3^+$ emission at Saturn (*O'Donoghue et al.*, 2013). The measured emission revealed significant latitudinal structure, with local extrema in one hemisphere being approximately mirrored at magnetically conjugate latitudes in the opposite hemisphere. Furthermore, those minima and maxima mapped to specific regions of Saturn's rings, implying a direct ring-atmosphere connection. These low- and mid-latitude $H_3^+$ emission structures have therefore been interpreted as representing the ionospheric signatures of "ring rain", a process wherein charged water products from Saturn's rings are transported along magnetic field lines into its atmosphere (*Connerney*, 2013). The spectral images obtained during those observations are shown in **Figure 8.5**; the $H_3^+$ intensities plotted as a function of planetocentric latitude are given in **Figure 8.6**.

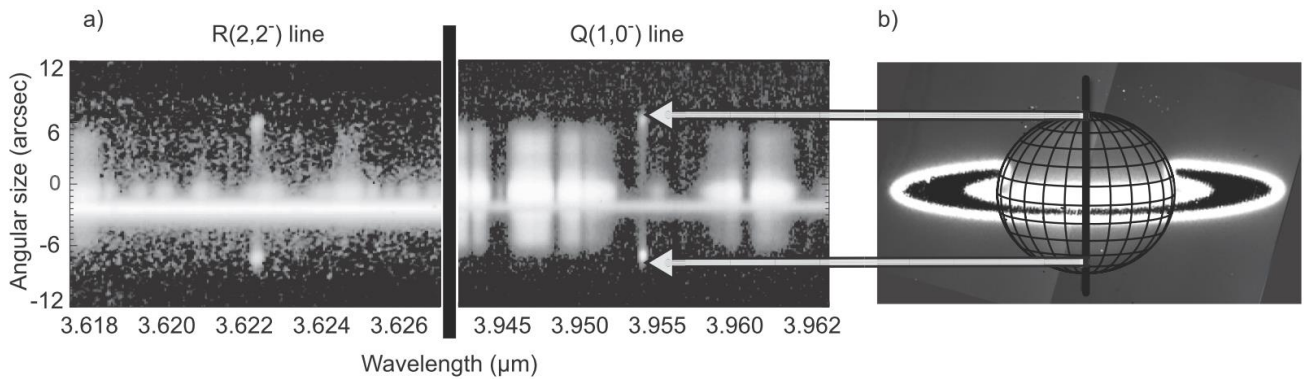

**Figure 8.5:** Two spectral regions are indicated in **(a)**, surrounding two bright rotational-vibrational $H_3^+$ emission lines: the R(2,2$^-$) line at 3.622 μm and the Q(1,0$^-$) line at 3.953 μm. The vertical axis corresponds to spatial position along the slit, which is oriented north-south along Saturn's rotational axis, as illustrated in **(b)**. Uniform reflection of sunlight by the rings can be seen as a bright horizontal patch of emission near -3 arcsec in (a). Remaining bright areas in the spectra are due to methane reflection. Saturn's aurorae are visible as the bright regions at either end of the R(2,2$^-$) and Q(1,0$^-$) lines, though emission is also clearly seen across the entire disk. From *O'Donoghue et al.* (2013).

---

While **Figures 8.5** and **8.6** clearly indicate there is some connection between Saturn's rings and its ionosphere, the nature of that connection is not definitively revealed by the measurements. The detection was too weak to derive column-integrated $H_3^+$ temperatures, as is commonly done for auroral $H_3^+$ measurements (e.g., *Melin et al.*, 2007), and consequently there is an ambiguity behind whether the emission structure is driven primarily by temperature or density variations. Based on earlier work, an enhanced influx of ring material is expected at latitudes that map to the inner edge of Saturn's B ring, roughly 44°N and 38°S planetocentric latitude (e.g., *Connerney and Waite*, 1984; *Connerney*, 1986; see also Section 8.2.1). These regions do correspond to local maxima in $H_3^+$ emission (**Figure 8.6**). Saturn's rings are predominantly water ice bodies (e.g., *Cuzzi et al.*, 2010, and references therein), and thus some sort of water product, such as sub-micrometer-sized grains with a high charge-to-mass ratio (*Connerney*, 2013), is the most likely ring material to precipitate into its atmosphere. More recent models of the ring ionosphere, prompted by the detection of $O_2^+$ and $O^+$ ions by Cassini during Saturn Orbital Insertion (*Tokar et al.*, 2005; *Waite et al.*, 2005), also predict a significant influx of those ions into Saturn's atmosphere (*Luhmann et al.*, 2006; *Tseng et al.*, 2010). A ring-atmosphere current system may lead to localized temperature enhancements due to Joule heating (*O'Donoghue et al.*, 2013), though preliminary calculations find the Joule heating rates to be negligible (*Crary*, 2014). Therefore, the observed $H_3^+$ emission maxima more likely represent either an enhancement in density in those regions or depletions in density in surrounding regions. An influx of water into Saturn's atmosphere has been invoked in the past in order to explain radio occultation observations, though the effect of such an influx on $H_3^+$ densities has been largely neglected until now. Hereafter, for simplicity, "water" is used as a catch-all term to describe ring-derived influxes into Saturn's atmosphere, as they are all expected to have the same primary effect: increased $H^+$ loss chemistry leading ultimately to a reduced electron density.

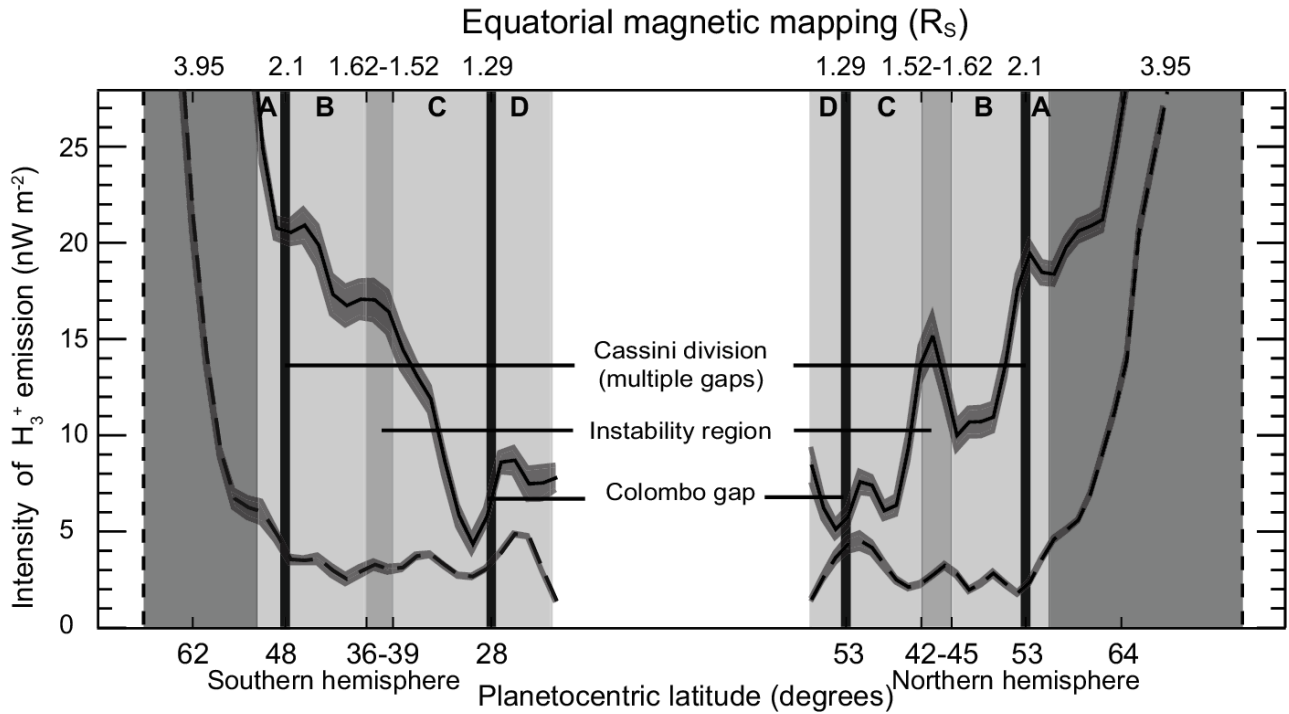

**Figure 8.6:** Intensity of $H_3^+$ infrared emission as a function of latitude along Saturn's noon meridian, based on the spectra of Figure 8.5. Planetocentric latitudes are indicated along the bottom axis, while the upper axis gives the equatorial radius each latitude element maps to along magnetic field lines. Gray shading envelopes surrounding the Q(1,0⁻) (solid curve) and R(2,2⁻) (dashed curve) spectral lines indicate the 1σ errors in intensity. From *O'Donoghue et al.* (2013).

## 8.3 Models

We have outlined the available observational constraints for Saturn's ionospheric parameters in Section 8.2. We now give an overview of the basic theory used to explain these measurements, and review the past modeling studies (Section 8.3.1). A more thorough description of past ionospheric modeling studies can be found in *Nagy et al.* (2009), and so we touch only briefly on that history here. Contemporary model-data comparisons are highlighted in order to give context to the current state of knowledge and to emphasize outstanding model-data discrepancies (Section 8.3.2).

### 8.3.1 Basic Theory: Chemistry, Ionization, and Temperature

Molecular hydrogen is the dominant constituent in Saturn's upper atmosphere, with atomic hydrogen becoming important at the higher altitudes. At non-auroral latitudes, the primary source of ionization is solar X-ray (0.1-10 nm) and EUV (10-110 nm) photons. As the dominant constituent, $H_2$ absorbs most of the incident radiation at those wavelengths, and so the vast majority of photo-produced ions are $H_2^+$. Maximum photoionization rates (i.e., overhead illumination for solar maximum photon fluxes) for $H_2^+$ and $H^+$ in Saturn's ionosphere are roughly 10 cm$^{-3}$ s$^{-1}$ and 1 cm$^{-3}$ s$^{-1}$, respectively (*Moore et al.*, 2004). Direct photoionization of methane also leads to an array of hydrocarbon ions; $CH_4^+$ is produced most rapidly, with a 2-3 cm$^{-3}$ s$^{-1}$ production rate, though these ions are produced much lower in the ionosphere, near the homopause, where their parent species are located (*Kim et al.*, 2014). While $H_2^+$ is the dominant species created by photoionization, it is quickly converted to $H_3^+$ via the charge-exchange reaction (*Theard and Huntress*, 1974; see also *Miller et al.*, 2006, and references therein):

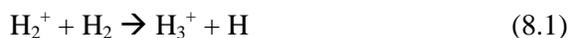

$$H_2^+ + H_2 \rightarrow H_3^+ + H \qquad (8.1)$$

Saturn's ionosphere is thus predicted to be dominated by a mix of $H_3^+$ and $H^+$ ions near the electron density peak and above, with an additional ledge of hydrocarbon ions below the peak, closer to the homopause (e.g., *Majeed and McConnell*, 1991; *Moses and Bass*, 2000; *Moore et al.*, 2004; *Kim et al.*, 2014). The lower ionosphere (e.g., ~2300 km and below for mid-latitudes) is expected to be in photochemical equilibrium; for such conditions the $H^+/H_3^+$ number density ratio is found to be proportional to electron density (*Moore et al.*, 2004). Dissociative recombination between $H_3^+$ and electrons – the dominant loss for $H_3^+$ ions – is rapid relative to the Saturn day. Typical $H_3^+$ lifetimes are ~10-15 minutes in Saturn's ionosphere (*Melin et al.*, 2011; *Tao et al.*, 2011), much shorter than Saturn's ~10 hour rotation period (see Chapter 5). Consequently, for conditions of reduced electron density, the $H^+/H_3^+$ ratio decreases, and the expected diurnal variation of the electron density in the main ionosphere increases (i.e., a smaller $H^+/H_3^+$ ratio means a larger fraction of short-lived $H_3^+$ ions).

Early models of Saturn's ionosphere (e.g., *McElroy*, 1973; *Capone et al.*, 1977) predicted a peak electron density an order of magnitude larger than subsequent Pioneer and Voyager radio occultation measurements revealed (*Kliore et al.*, 1980; *Lindal et al.*, 1985). The only chemical loss included for $H^+$ in these early models was radiative recombination, an extremely slow process. Therefore, in order to reduce modeled electron densities to better reproduce the observed values, a method of converting $H^+$ to a short-lived ion was required. The two methods considered by most subsequent models are charge exchange between $H^+$ and vibrationally excited $H_2$, and charge exchange with water introduced into the atmosphere from Saturn's rings and/or icy moons.

*Vibrationally Excited $H_2$*

As was recognized early on (*McElroy*, 1973), the charge-exchange reaction

$$H^+ + H_2(v≥4) \rightarrow H_2^+ + H \quad (8.2)$$

is exothermic only when $H_2$ is in the 4$^{th}$ or higher vibrational level. While there has historically been some uncertainty regarding the R8.2 reaction rate, it has generally been assumed to be of the order 1-2x10$^{-9}$ cm$^3$ s$^{-1}$ (e.g. *McConnell et al.*, 1982; *Cravens*, 1987). Recent extrapolation of work by the plasma fusion quantum theory community (e.g., *Ichihara et al.*, 2000; *Krstić*, 2002) has been used to refine the estimated R8.2 reaction rate (at 600 K) only slightly, to 0.6-1.3x10$^{-9}$ cm$^3$ s$^{-1}$ (*Huestis*, 2008). Of far greater uncertainty is the population of non-LTE vibrationally excited $H_2$, where LTE refers to local thermodynamic equilibrium. (If the $H_2$ vibrational distribution were in LTE at 1000 K, fewer than 10$^{-10}$ molecules would be in the $v≥4$ state, far too low a value to significantly impact $H^+$ densities.) Molecular hydrogen in giant planet upper atmospheres can be vibrationally excited by electron impact, by solar and electron excitation of the $H_2$ Lyman and Werner bands, which then fluoresce to vibrationally excited levels in the ground state, and by dissociative recombination of $H_3^+$ (e.g., *Waite et al.*, 1983; *Cravens*, 1987; *Majeed et al.*, 1991). Major vibrational loss processes include de-excitation through collisions with H and $H_2$, reactions with $H^+$ (i.e., R8.2), and redistribution of vibrational quanta among molecular levels through vibrational-vibrational (V-V) collisions and in altitude through diffusion.

The first detailed model for $H_2(v)$ in the outer planets was presented by *Cravens* (1987) for Jupiter. *Majeed et al.* (1991) added solar fluorescence in their low-latitude solar input $H_2(v)$ calculations for Jupiter and Saturn, finding it to be a dominant source of vibrational excitation. Both *Cravens* (1987) and *Majeed et al.* (1991) found modest enhancements in $H_2(v≥4)$ populations, leading to reductions in calculated electron densities that were insufficient to reproduce the observed electron density profiles. In general, calculations of $H_2(v≥4)$ appear to fall short of the enhanced vibrational populations required to bring modeled and observed electron densities into agreement, possibly due to uncertainties in the calculations (such as rate coefficients or source mechanisms; *Majeed et al.*, 1991), or possibly due to other processes acting to reduce $H^+$ densities. More recently, *Hallett et al.* (2005a) developed a new rotational-level hydrogen physical chemistry model, and subsequently applied it to Uranus (*Hallett et al.*, 2005b).

Subsequent model reproductions of electron density profiles from radio occultation observations have typically modified $H_2(v≥4)$ populations freely or have used scaled versions of the *Majeed et al.* (1991) calculations (e.g., *Majeed and McConnell*, 1996; *Moses and Bass*, 2000; *Moore et al.*, 2006). In other words, partly due to uncertainties in direct $H_2(v≥4)$ calculations, and partly due to attention being focused on other details of the Saturn ionosphere, contemporary models have predominantly treated the population of vibrationally excited $H_2$ as a free parameter. Reaction R8.2 directly reduces $H^+$ ion densities by converting $H^+$ to short-lived molecular ions, thereby indirectly reducing the net electron density and reducing the dominant chemical loss for $H_3^+$ – dissociative recombination with electrons. Due to the long lifetime of $H^+$, R8.2 can also act as an additional source of $H_2^+$ (and $H_3^+$ via R8.1), even in regions absent of ionizing radiation or precipitating energetic particles. Therefore, as it affects all of the major chemistry, R8.2 and the true population of vibrationally excited $H_2$ remain major points of uncertainty for ionospheric calculations. There are no direct observational constraints published at present.

*Water in Saturn's Ionosphere*

A second likely method of converting $H^+$ ions into short-lived molecular ions, thereby depleting the calculated electron densities, begins with an influx of water into Saturn's atmosphere. Possible external water sources include micrometeorites as well as Saturn's rings and icy satellites. This ionospheric quenching chain was first postulated by *Shimizu* (1980), implemented by *Chen* (1983), and treated comprehensively by *Connerney and Waite* (1984).

$$H^+ + H_2O \rightarrow H_2O^+ + H \quad (8.3)$$

$$H_2O^+ + H_2 \rightarrow H_3O^+ + H \quad (8.4)$$

$$H_3O^+ + e \rightarrow \begin{cases} H_2O + H & (8.5a) \\ H_2 + OH & (8.5b) \\ H + H + OH & (8.5c) \end{cases}$$

As noted by *Connerney and Waite* (1984), any OH in the system (such as that from R8.5) has a short lifetime due to its reaction with $H_2$, producing $H_2O$ and H. Similarly, any ionized water products, such as $O_2^+$, dissociatively recombine with electrons extremely rapidly – roughly three times faster than $H_3^+$ in Saturn's ionosphere – leading to a chain of photochemical reactions that produce primarily OH (via $O + H_2$) and $H_2O$ (via $OH + H_2$) in the thermosphere and lower atmosphere (e.g. *Moses and Bass*, 2000; *Moses et al.*, 2000). Hence, while the exact form of exogenous influx may not always be pure $H_2O$, the ionospheric chemical effects are similar.

The various reaction rates for water chemistry in outer planet upper atmospheres are relatively well known (e.g., *Moses and Bass*, 2000; *Moses et al.*, 2000). Of far greater uncertainty is the magnitude of water influx at Saturn, as well as its spatial and temporal distribution and variability. While a number of modeling studies have derived a range of water influxes indirectly as a means of reproducing the electron density profiles from radio occultations (e.g., *Connerney and Waite*, 1984; *Majeed and McConnell*, 1991, 1996; *Moore et al.*, 2006, 2010), directly constraining the influxes observationally has proven more difficult. The first unambiguous detection of water in Saturn's upper atmosphere came from the Infrared Space Observatory (ISO; *Feuchtgruber et al.*, 1997), which measured an $H_2O$ column abundance of $(0.8-1.7) \times 10^{15}$ cm$^{-2}$ and was used to derive a global water influx of $\sim 1.5 \times 10^6$ $H_2O$ molecules cm$^{-2}$ s$^{-1}$ (*Moses et al.*, 2000). Subsequent studies based on Submillimeter Wave Astronomy Satellite and Herschel Space Observatory measurements found global influx values within a factor of 4 of the *Moses et al.* result (*Bergin et al.*, 2000; *Hartogh et al.*, 2011). Despite predictions of strong latitudinal variations in water influx (e.g., *Connerney*, 1986), no observational confirmation of such variations has yet been published. At present, there are ambiguous detections of latitudinally varying water volume mixing ratios in the ultraviolet (e.g., a 2σ detection of $2.70 \times 10^{16}$ cm$^{-2}$ at 33°S planetocentric latitude: *Prangé et al.*, 2006) as well as preliminary indications of larger equatorial water densities from Cassini Composite InfraRed Spectrometer (CIRS) observations (*Bjoraker et al.*, 2010) and from further Herschel observations (*Cavalié et al.*, 2014).

A number of different categories of theoretical studies support the preliminary observational results that favor a latitudinal variation of water influx at Saturn. The first category includes ring modeling studies focused on the erosion of Saturn's rings (e.g., *Northrop and Hill*, 1982, 1983; *Ip*, 1983; *Northrop and Connerney*, 1987). One of the outcomes of such studies is the demonstration that small negatively charged ring grains inside of a marginal stability radius of 1.525 $R_S$ can be lost to Saturn's atmosphere along magnetic field lines. A recent related study also explores the evolution of positively charged ring grains, and finds that they are sometimes deposited in Saturn's equatorial atmosphere (*Liu and Ip*, 2014). A second category of studies is focused on the ring atmosphere and ionosphere, and likewise predicts a precipitation of ring ions into Saturn's atmosphere (e.g., *Luhmann et al.*, 2006; *Tseng et al.*, 2010). In general, all of the ring models predict an asymmetry in the particle influx due to Saturn's slightly offset magnetic dipole, with stronger influxes expected in the southern hemisphere. Finally, a third main category of modeling studies that predict a latitudinal variation of water influx at Saturn are those that track the water vapor ejected from Enceladus' plumes (*Porco et al.*, 2006). These models estimate that approximately 10%, 7%, 3%, and 6%, respectively, of the Enceladus water is lost to Saturn's atmosphere (*Jurac and Richardson*, 2007; *Cassidy and Johnson*, 2010; *Hartogh et al.*, 2011; *Fleshman et al.*, 2012). Latitude variations of Saturn water influx from the Enceladus models vary, though in general a stronger influx is predicted at low-latitudes. Finally, temporal (and possibly seasonal) variability is also expected for both ring and Enceladus sources of water, though it is not well constrained at present.

*Primary and Secondary Ionization*

Ionizing radiation at the giant planets comes primarily in two forms: solar photons and energetic particles. Electrons released during ionization are usually suprathermal. These suprathermal electrons – referred to as photoelectrons in the case of photoionization and secondary electrons in the case of particle impact ionization – possess enough energy to excite, dissociate, and further ionize the neutral atmosphere as well as to heat the plasma. On the one hand, photoionization production rates follow from a fairly straightforward application of the Lambert-Beer Law, at least assuming that the neutral atmospheric densities, incident solar fluxes, and photoabsorption and photoionization cross sections are known. Conversely, in order to accurately track the transport, energy degradation, and angular redistribution of suprathermal electrons, a kinetic approach needs to be applied, such as solution to the Boltzmann equation through a multi-stream approach (*Perry et al.*, 1999; *Moore et al.*, 2008; *Galand et al.*, 2009; *Gustin et al.*, 2009) or a Monte Carlo approach (e.g., *Tao et al.*, 2011). While ionization due to photoelectrons and secondary electrons is usually included in the case of Venus, Earth, Mars, Jupiter, and Titan (e.g., *Kim et al.*, 1989; *Kim and Fox*, 1994; *Schlesier and Buonsanto*, 1999; *Cravens et al.*, 2004; *Fox*, 2004, 2007; *Fox and Yeager*, 2006; *Galand et al.*, 2006; *Matta et al.*, 2014), it is this added complexity that has prevented secondary ionization from being treated in the majority of Saturn ionospheric models.

Higher energy photons produce higher energy photoelectrons, and therefore lead to more secondary ionization. In general photoabsorption cross sections are smallest for high energy photons and increase nearly monotonically over their ionization range. There are certainly exceptions to this generality (e.g., methane near 4 nm), however it holds true for $H_2$ for photons >1 nm – the dominant absorber in the outer planets – meaning that more energetic photons are typically absorbed lower in the atmosphere at Saturn (*Moses and Bass*, 2000; *Galand et al.*, 2009). Secondary ionization due to solar illumination, therefore, primarily increases the ion production rate in the lower ionosphere.

In the auroral regions, precipitating particles interact with the ambient atmosphere via collisions, leading to excitation, ionization and heating. About half of all inelastic collisions between precipitating energetic electrons and Saturn's upper atmosphere result in the ionization of $H_2$ that is initially in the electronic ground state $(X^1\Sigma_g^+)$, producing both $H_2^+$ and secondary electrons $e_s$:

$$e + H_2 \rightarrow H_2^+ + e_s + e \qquad (8.6)$$

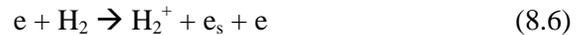

Since this process does not remove the energetic electrons, the precipitating electrons and their secondary products undergo further inelastic collisions, producing additional ionization, excitation, and dissociation in the atmosphere, as well as heating of thermal, ionospheric electrons. These energetic particles lose kinetic energy with each collision until they are finally thermalized with the surrounding atmosphere. Atmospheric effects due to precipitating energetic electrons (e.g., *Galand et al.*, 2011; *Tao et al.*, 2011), such as ionization and heating, are discussed in more detail in Chapters 7 and 9.

There are two studies that treat secondary ionization comprehensively in Saturn's non-auroral ionosphere. The first, *Galand et al.* (2009), solves the Boltzmann equation for suprathermal electrons using a multistream transport model based on the solution proposed by *Lummerzheim et al.* (1989) for terrestrial applications. A simple parameterization of secondary ionization rates based on the *Galand et al.* (2009) results appears in *Moore et al.* (2009), accurate over a range of solar/seasonal conditions and latitudes. A number of related modeling studies include either the *Galand et al.* (2009) results or the *Moore et al.* (2009) parameterization (e.g., *Moore et al.*, 2010; *Barrow and Matcheva*, 2013). The second study to calculate secondary ionization rates at Saturn directly, *Kim et al.* (2014), assumes that photoelectrons deposit their energy locally using a simple method described by *Dalgarno and Lejeune* (1971). A similar approach has been used for Jupiter (e.g., *Kim and Fox*, 1994). Both the *Galand et al.* (2009) and the *Kim et al.* (2014) studies are for mid-latitude, and both find that the secondary ionization production rates are roughly comparable to primary photoionization rates just below 1000 km altitude (i.e., for pressures greater than ~$10^{-5}$ mbar). At lower altitudes (higher pressures) secondary ionization production rates are dominant by as much as an order of magnitude. The effect on calculated ion and electron densities is also altitude-dependent: electron densities are increased by roughly 30% at the peak and by up to an order of magnitude at lower altitudes (*Galand et al.*, 2009). The impact of secondary ionization on Saturn ionospheric electron densities is illustrated in **Figure 8.7**, which shows the ratio of calculated electron densities between simulations that do and do not include the extra production term. It is clear from **Figure 8.7** that models of Saturn's ionosphere that do not account for secondary ionization in some form will significantly underestimate electron densities in Saturn's lower ionosphere.

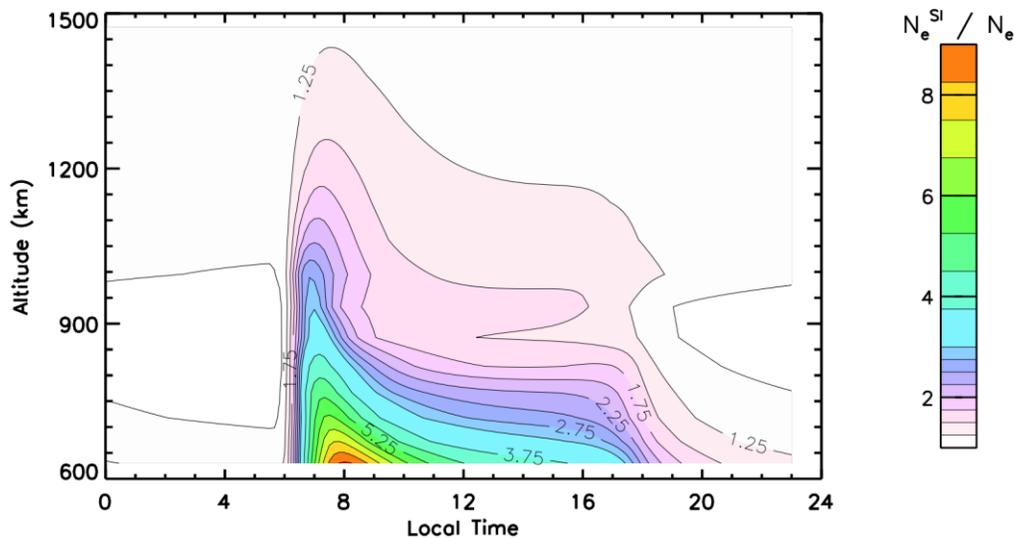

**Figure 8.7:** Contour plot of the electron density ratio between a simulation that includes secondary ionization ($N_e^{SI}$) and a run that ignores secondary ionization ($N_e$). Both simulations are for solar minimum conditions at mid-latitude. From *Galand et al.* (2009).

---

*Hydrocarbon ions*

Most of the preceding discussion has focused on $H^+$ and $H_3^+$, as those are the ions predicted to be dominant throughout most of Saturn's ionosphere. There is, however, an additional predicted ledge of low-altitude ionization, thought to be dominated by hydrocarbon (and possibly metallic) ions, just above the homopause. Many models treat the hydrocarbon layer as an ionospheric sink, if they consider it at all, as methane readily charge-exchanges with $H^+$ and $H_3^+$, leading to a relatively short-lived molecular hydrocarbon ions (e.g., *Moore et al.*, 2008, and related studies). Such a treatment can lead to fairly accurate electron densities within the hydrocarbon layer, at least when compared with more comprehensive models, but the resulting hydrocarbon ion composition is incorrect (e.g., *Moses and Bass*, 2000; *Moore et al.*, 2008).

There are two main complications that models must address in order to treat hydrocarbon ions comprehensively. First, there are numerous hydrocarbon ions and a significantly more complex series of reactions to consider. Depending on the modeling approach, this may only hinder results by requiring a larger table of ions and reactions to be inserted, though even in that case many of the reaction rates are unknown or untested in the laboratory. The two models that treat the hydrocarbon layer at Saturn most comprehensively are *Moses and Bass* (2000) and *Kim et al.* (2014). *Moses and Bass* consider 109 different ion species with 845 reactions, while *Kim et al.* track 53 ions using 749 reactions. Note that models developed for Titan's rich high-molecular-weight hydrocarbon atmosphere include an even more complete list of reactions and ions (e.g.,*Waite et al.*, 2010; *Vuitton et al.*, 2015). The second complication that needs to be addressed for accurate calculations of hydrocarbon ion densities is that high resolution $H_2$ photoabsorption cross sections (of the order of $10^{-4}$ nm) are required between ~80 nm (the $H_2$ ionization threshold) and 111.6 nm. Photons across this wavelength range, in which $H_2$ absorbs in discrete transitions – mostly in the Lyman, the Werner, and the Rydberg band systems – possess enough energy to ionize atomic hydrogen as well as methane, and $H_2$ photoabsorption cross sections can differ by six orders of magnitude over very short wavelength scales. Calculations that use low resolution cross sections will absorb these photons higher in the atmosphere, on average, before they can ionize methane and other hydrocarbon neutrals near the homopause; such models consequently under predict photoionization rates within the hydrocarbon layer. The only study so far to include high resolution $H_2$ photoabsorption cross sections at Saturn is *Kim et al.* (2014). Steady-state ion density profiles from the *Kim et al.* (2014) calculations are shown in **Figure 8.8**, where the high resolution (0.0001 nm) and low resolution (0.1 nm) model densities are indicated by solid and dotted curves, respectively.

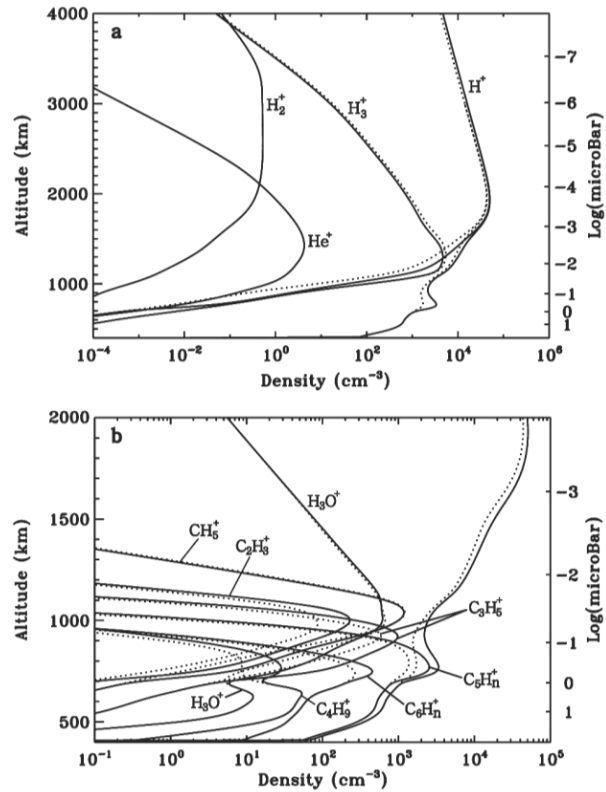

**Figure 8.8:** Steady state density profiles of **(a)** (non-hydrocarbon) H- and He-bearing ions, and **(b)** major hydrocarbon and oxygen-bearing ions, as well as total electron density (thick solid line, both panels). Solid and dotted curves represent densities from the high spectral resolution (0.0001 nm) and low spectral resolution (0.1 nm) models, respectively. Note the different altitude and density scales in each panel. From *Kim et al.* (2014).

*Kim et al.* (2014) find photoionization rates from the high resolution model are much larger than those from the low resolution model for some ion species. For example, at low altitude, near 830 km (0.2 µbar), *Kim et al.* (2014) find the photo production rate of $H^+$ (from H) is enhanced by a factor of ~25. At a slightly lower altitude, the production rate of $CH_4^+$ is larger by a factor of ~22. Due to rapid photochemical loss rates, however, these differences are not as dramatic for the calculated ion densities. The high resolution model ion densities (solid lines) are more moderately enhanced relative to the low resolution model ion densities (dotted lines) in **Figure 8.8**: the sum of the ion densities at the hydrocarbon peak is ~3200 cm$^{-3}$ and ~1800 cm$^{-3}$ for the high resolution and low resolution models, respectively.

*Plasma temperatures*

Ion, electron, and neutral temperatures are expected to deviate in the upper atmosphere of Saturn, though no in situ measurements have yet been made. Photoelectron (and secondary electron) interactions with the ambient plasma are likely the dominant source of heating in the non-auroral ionosphere, and therefore plasma temperature model calculations require an accurate treatment of the transport, energy degradation, and angular redistribution of suprathermal electrons. Plasma temperatures affect ionospheric model calculations primarily by altering chemical reaction and ambipolar diffusion rates.

There are two model calculations for plasma temperatures at Saturn available in the refereed literature, one at high latitudes (*Glocer et al.*, 2007), and one at mid latitudes (*Moore et al.*, 2008). (Two other previous studies are from Ph.D. dissertations: *Waite*, 1981; *Frey*, 1997.)

*Glocer et al.* (2007) use a 1-D multi-fluid model to study the polar wind at Saturn, starting from below the ionospheric peak and extending to an altitude of 1 $R_S$, yielding densities, fluxes, and temperatures for $H^+$ and $H_3^+$. They find peak ion temperatures of roughly 2000-3200 K for $H^+$ and 1300-2600 K for $H_3^+$ – well above the main ionosphere sampled by remote auroral observations. *Moore et al.* (2008) self-consistently coupled a 1-D ionospheric model that solves the ion continuity, momentum, and energy equations with a suprathermal electron transport code adapted to Saturn (*Galand et al.*, 2009). Their calculations specified a fixed neutral background based on results from 3-D global circulation calculations that reproduced observed thermospheric temperatures (*Müller-Wodarg et al.*, 2006). *Moore et al.* (2008) found only relatively modest electron temperature enhancements during the Saturn day, calculating peak values of ~500-560 K (~80-140 K above the neutral temperature). Ion temperatures were somewhat smaller, reaching ~480 K at the topside during the day while remaining nearly equal to the neutral temperature at altitudes near and below the electron density peak. Both ions and electrons cooled to the neutral temperature within an hour after sunset. A parameterization of the thermal electron heating rate based on primary photoionization rates was also developed (*Moore et al.*, 2008) and then demonstrated to work for a wide variety of seasonal/solar conditions and latitudes (*Moore et al.*, 2009).

Plasma temperatures can also be estimated from the topside scale heights of observed electron densities, though there are a number of uncertainties associated with such an estimate. For example, the ion composition has not been measured, and there may be small altitude gradients in temperature. Both of these unknowns can lead to ambiguous results. Nonetheless, as most Saturn models predict $H^+$ as the dominant topside ion, especially at dawn, *Nagy et al.* (2006) assumed $H^+$ was the major topside ion and neglected possible temperature gradients in order to arrive at an estimate of 625 K based on analyzing the average low-latitude dawn radio occultation profile above 2500 km altitude. By applying the same assumptions, and by considering that dusk temperatures should be at least as large as dawn temperatures, *Nagy et al.* (2006) further arrive at the implication that the dusk topside might be 72% $H_3^+$, or 7.7% $H_3O^+$, or some other appropriate combination of ion fractions.

### 8.3.2 Model-Data Comparisons

There are five major categories of observational constraints that must be explained by models: (1) peak electron density and altitude, (2) dawn/dusk electron density asymmetry, (3) latitudinal variations in $N_{MAX}$ and TEC, (4) diurnal variation of $N_{MAX}$, and (5) latitudinal $H_3^+$ structure. While a number of these observational constraints are closely related, it is illustrative to review model-data comparisons for each separately in order to highlight the key parameters that drive each of the observables. Models typically attempt to reproduce two or more of the observables simultaneously, though this is often not possible due to the different solar, seasonal and latitudinal conditions of the multi-instrumental observations.

*Electron Density Altitude Structure*

Peak electron densities in Saturn's ionosphere ($N_{MAX}$), based on Cassini radio occultations, range from ~$0.3 \times 10^3$-$2.6 \times 10^4$ cm$^{-3}$ at dawn and ~$3 \times 10^3$-$1.9 \times 10^4$ at dusk. The altitude of the electron density peak, $h_{MAX}$, has been observed at altitudes between 1100 km and 3200 km (Section 8.2.1). In general, both $h_{MAX}$ and $N_{MAX}$ are smallest near Saturn's equator, and increase with latitude, though there is significant scatter about this average behavior (*Kliore et al.*, 2014).

The primary focus of early Saturn ionospheric models following the observations of the Pioneer and Voyager spacecraft was on reducing modeled electron densities to the measured ~$10^4$ cm$^{-3}$ peak value. As can be seen from Figure 8.1, there is very rarely a classic Chapman-type smooth electron density profile at Saturn, and the maximum electron density is at times located in a narrow low-altitude layer rather than what might be called the "main peak". For these reasons, models have generally focused on understanding the average trends revealed by observations rather than on reproducing exact electron density altitude profiles. Individual profiles have also been reproduced in the past, typically by varying a number of free parameters (e.g., *Majeed and McConnell*, 1991): the population of vibrationally excited $H_2$, the external water influx, and the vertical plasma drift. The first two of these parameters have been discussed already (Section 8.3.1); vertical plasma drift can arise from neutral winds (e.g., meridional winds driving plasma up or down magnetic field lines) and electric fields (e.g., **E** x **B** drifts driven by zonal electric fields; *Kelley*, 2009; *Schunk and Nagy*, 2009). Whereas the primary effect of an enhanced population of vibrationally excited $H_2$ or water influx is to reduce the modeled electron densities, the peak altitude is also slightly increased as these reactions are more effective at lower altitudes (e.g., *Majeed and McConnell*, 1991; *Moses and Bass*, 2000; *Moore et al.*, 2004). Similarly, the primary effect of a vertical plasma drift is to shift $h_{MAX}$ up or down in altitude, though $N_{MAX}$ can also be affected if the plasma is moved into a different chemical or dynamical regime.

Though it is possible to construct a model reproduction of most of the observed electron density altitude profiles by exploring unknown parameters (though likely not all of them), the derived parameters vary significantly from observation to observation. No single set of water influxes, vibrationally excited $H_2$ populations, and enforced vertical plasma drifts can reproduce all of the observed radio occultation electron densities simultaneously, possibly indicating spatial and temporal variations in these parameters.

Beyond simply comparing to $N_{MAX}$ and $h_{MAX}$, modelers have also attempted to understand the narrow layers of electron density frequently observed (Figures 8.1-8.2). Vertically varying horizontal winds, such as might occur from atmospheric gravity waves, can cause alternating compression and extension of plasma with altitude, thus creating ionospheric layers (*Kelley*, 2009). Such layering is frequently observed in the terrestrial E region and the lower F region. This mechanism is especially effective for long lived ions, such as the metallic ions introduced from the ablation of micrometeoroids. *Moses and Bass* (2000) were able to demonstrate the plausibility of such a layering process in Saturn's ionosphere by introducing a modest oscillatory vertical plasma drift as well as a Mg influx of $1.3 \times 10^5$ cm$^{-2}$ s$^{-1}$ focused in the 790-1290 km region. *Kim et al.* (2014) also suggest that such layers may result from photochemistry driven by high resolution H$_2$ photoabsorption cross sections.

Based on a wavelet analysis of 31 Cassini radio occultations (*Nagy et al.*, 2006; *Kliore et al.*, 2009), *Matcheva and Barrow* (2012) were able to detect several discrete scales of variability in Saturn's electron density profiles. Furthermore, by applying a gravity wave propagation model to Saturn's upper atmosphere, they also demonstrated that the observed features were consistent with gravity waves being present in the lower ionosphere, causing layering of the ions and electrons. In a continuation of that study, wherein a 2-D, non-linear, time-dependent model of the interaction of atmospheric gravity waves with ionospheric ions was applied to Saturn's upper atmosphere, *Barrow and Matcheva* (2013) were able to reproduce the structure of two Cassini radio occultations, the S08 entry and the S56 exit. **Figure 8.9** presents the model-data comparison for the S56 occultation.

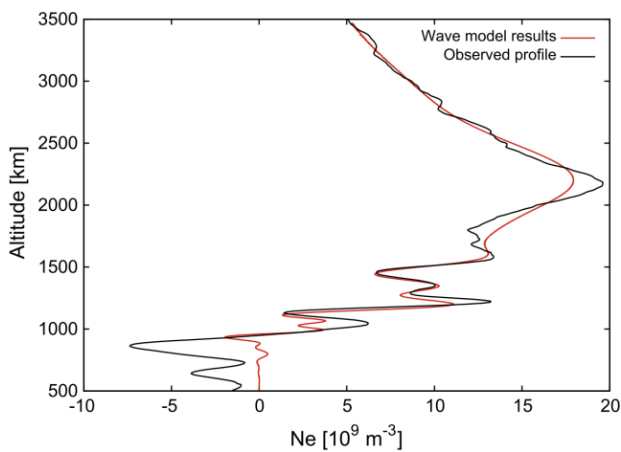

**Figure 8.9:** Comparison between the S56 exit observation (black line) and a model electron density profile (red line). The ionospheric model illustrates the effect of three small-amplitude gravity waves superimposed on the background electron density profile. From *Barrow and Matcheva* (2013).

Extreme "bite-outs", or localized electron density depletions, frequently seen in Saturn radio occultations, can also be produced from atmospheric gravity waves, as discussed above (see also Figure 6 of *Kliore et al.*, 2009). Another possible generation mechanism for ionospheric depletions is a time-dependent water influx. *Moore and Mendillo* (2007) were able to reproduce the observed S7 bite-out by increasing the background water influx (of $5 \times 10^6$ cm$^{-2}$ s$^{-1}$) by a factor of 50 for ~27 minutes, after which it returned to its initial value. As the resulting bulge of water density diffuses downward through the thermosphere, it undergoes charge exchange reactions with H$^+$, leading to a localized reduction in electron density. While a number of Cassini radio occultations contain depletions similar to the *Moore and Mendillo* (2007) results, the magnitude and frequency of possible water influx variations are not known at present. Therefore this possibility remains unconfirmed. Nevertheless, atmospheric gravity waves have difficulty producing ionospheric structure at high altitude due to wave dissipation lower in the thermosphere (*Matcheva and Barrow*, 2012), and as some electron density depletions have been observed above 2000 km (*Kliore et al.*, 2009), an alternative mechanism for generating such structures, such as a time-variable water influx, may yet be required to reproduce some observed electron density altitude profiles.

*Dawn/Dusk Electron Density Asymmetry*

The first dozen radio occultations obtained by Cassini revealed a dawn/dusk asymmetry in Saturn's low-latitude ionosphere (*Nagy et al.*, 2006). On average the peak densities are lower and the peak altitudes higher at dawn than at dusk. A recently developed global circulation model (GCM) of Saturn's upper atmosphere, called STIM (the Saturn Thermosphere Ionosphere Model), was leveraged in order to study this new observational constraint. In comparing with the first dozen Cassini equatorial radio occultations, 1-D ionospheric calculations were performed (*Moore et al.*, 2006) using a 3-D thermosphere that reproduced observed upper atmospheric temperatures using a combination of auroral Joule heating and low-latitude wave heating (*Müller-Wodarg et al.*, 2006). *Moore et al.* (2006) found that the average dawn and dusk equatorial electron density profiles were best reproduced by model simulations that considered a water influx at the top of the atmosphere of $5 \times 10^6$ cm$^{-2}$ s$^{-1}$. Electron density comparisons are shown in **Figure 8.10**. This water influx was roughly a factor of three larger than the globally averaged influx derived from ISO observations, $1.5 \times 10^6$ cm$^{-2}$ s$^{-1}$ (*Moses et al.*, 2000), though such a difference is likely not drastic enough to conflict with the spatial constraints evaluated by *Moses et al.* (2000). The fact that the *Moore et al.* (2006) water influx is smaller than the favored value derived from previous comparisons with the Voyager 2 exit occultation, $2.2 \times 10^7$ cm$^{-2}$ s$^{-1}$ (*Majeed and McConnell*, 1991), can be explained by the fact that the *Majeed and McConnell* (1991) water influx was derived for a scenario in which electron loss due to reactions between protons and vibrationally excited $H_2$ was not considered. By including this reaction in addition to charge exchange with water, the magnitude of each required rate is reduced: *Moore et al.* (2006) favored a population of vibrationally excited $H_2$ that was 25% the nominal value considered by *Majeed and McConnell* (1991).

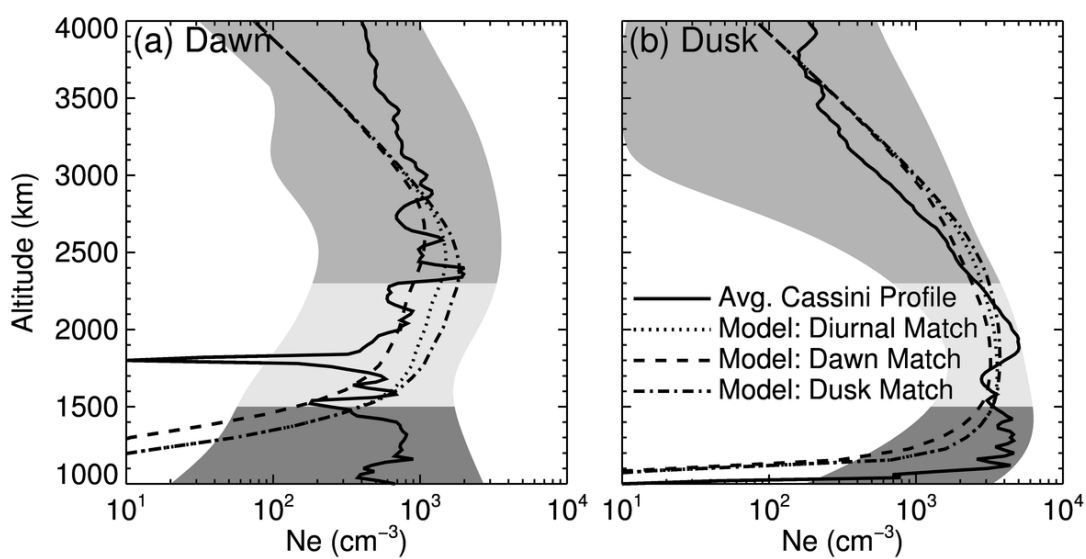

**Figure 8.10:** The average **(a)** dawn and **(b)** dusk Cassini electron density profiles (*Nagy et al.*, 2006) along with model comparisons. Dotted lines represent model results that best match the observations at both dusk and dawn using a full diurnal calculation with a single set of parameters, whereas dashed and dot-dashed lines give results best matched to the average dawn or dusk profile, respectively. The width of the shaded regions corresponds to the full range of electron densities observed by Cassini, and the degree of shading represents three distinct ionospheric altitude regimes, from top to bottom: diffusive regime (dominated by $H^+$), photochemical regime (dominated by $H^+$ and $H_3^+$), and hydrocarbon/metallic ion regime (where these ions begin to dominate). From *Moore et al.* (2006).

The diurnal variation in peak electron density implied by the model reproductions of average dawn and dusk Cassini occultations was modest, less than a factor of 6, as illustrated in **Figure 8.11.** A similar set of parameters was able to reproduce the dawn and dusk $N_{MAX}$ values measured by Cassini for a majority of the twelve equatorial occultations, with the notable exceptions being S11x, S9x, and S12n/x (*Moore et al.*, 2006).

The dawn/dusk asymmetry is primarily due to the presence of both atomic ($H^+$) and molecular ($H_3^+$) ions at the electron density peak. At dusk the solar source of ionization has only just shut off, so both $H^+$ and $H_3^+$ are still present and contribute to the electron density peak. During the ~5 hours of darkness on the nightside, however, most of the $H_3^+$ ions have dissociatively recombined with electrons, resulting in a reduced electron density at dawn. This effect is similar to another well-known atomic and molecular ion region, the terrestrial F-layer. Modeled ion and electron altitude profiles at dawn, noon, and dusk, are presented in **Figure 8.12**, demonstrating the loss of $H_3^+$ ions during the Saturn night (*Moore et al.*, 2008). Earlier Saturn ionospheric calculations produce the same sort of variations in $H_3^+$, $H^+$, and $N_e$, though for different seasonal and solar conditions (e.g., Figure 6 of *Majeed and McConnell*, 1996; Figure 13 of *Moses and Bass*, 2000).

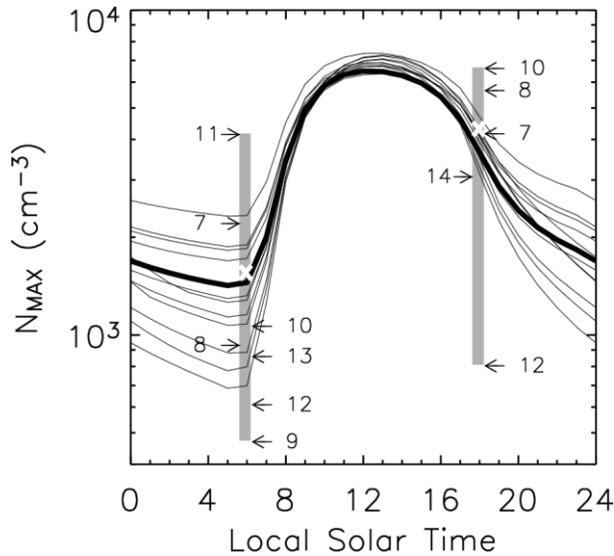

**Figure 8.11:** Plot of modeled local time variations of $N_{MAX}$. Each solid line represents the best diurnal match for one set of combinations of the two unknown chemical losses (due to vibrationally excited $H_2$ and water influx). Gray shaded rectangles identify the ranges in local solar time and $N_{MAX}$ from the Cassini radio occultations (*Nagy et al.*, 2006). Numbers and arrows mark the individual occultation values and "x" marks the averaged dawn and dusk observations. From *Moore et al.* (2006).

*Latitudinal Variations in $N_{MAX}$ and TEC*

Analysis of 19 additional Cassini radio occultations mostly at mid- and high-latitudes (*Kliore et al.*, 2009), in addition to the 12 previous equatorial occultations (*Nagy et al.*, 2006), revealed a clear latitudinal trend in electron density: on average, electron densities were smallest at Saturn's equator and increased with latitude. This trend was recently strengthened by the addition of 28 more radio occultations across all latitudes (**Figure 8.3**; *Kliore et al.*, 2014). The sub-solar point was near Saturn's equator for a majority of the occultations, and therefore the presence of a minimum in electron density in the region of peak photoionization was most likely indicative of a latitudinal dependence in the chemical loss rates.

*Moore et al.* (2010) explored the possibility of a latitudinally varying chemical loss using 2-D ionospheric calculations combined with a fixed 3-D thermospheric background appropriate for the average seasonal and solar conditions for the first 31 Cassini radio occultations (*Kliore et al.*, 2009). They evaluated a Gaussian water influx centered at Saturn's equator with a peak value of $(0.1-1) \times 10^7$ cm$^{-2}$ s$^{-1}$ and a full width half maximum (FWHM) between $2°$-$180°$ (in addition to a latitudinally invariant water influx). Six populations of vibrationally excited $H_2$ were also considered, based on modifications to the *Majeed et al.* (1991) values by a factor of 0-2. The observed latitudinal trend in electron density was best reproduced for a Gaussian water influx centered at the equator with a FWHM of ~23.5°, illustrated in the left panel of **Figure 8.13**. The globally averaged water influx from such a profile is $1.1 \times 10^6$ cm$^{-2}$ s$^{-1}$, slightly less than the $1.5 \times 10^6$ cm$^{-2}$ s$^{-1}$ inferred from ISO measurements (*Moses et al.*, 2000; *Moore et al.*, 2015).

*Müller-Wodarg et al.* (2012) applied the *Moore et al.* (2010) water distribution in a subsequent study focused on magnetosphere-atmosphere coupling at high latitudes using the full 3-D STIM. **Figure 8.13** illustrates the imposed water influx and the latitude variation of both modeled and observed electron densities with latitude, demonstrating that a combination of latitudinally varying water influxes and seasonal trends can explain the majority of $N_{MAX}$ variations from radio occultation observations.

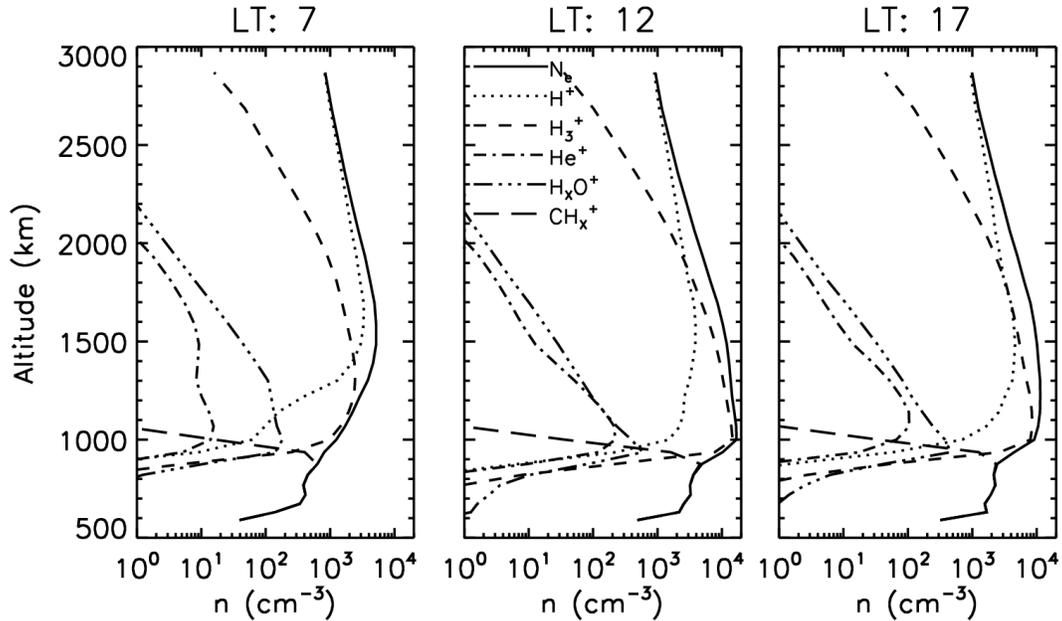

**Figure 8.12:** Modeled ion and electron densities at **(left)** dawn, **(middle)** noon, and **(right)** dusk. From *Moore et al.* (2008).

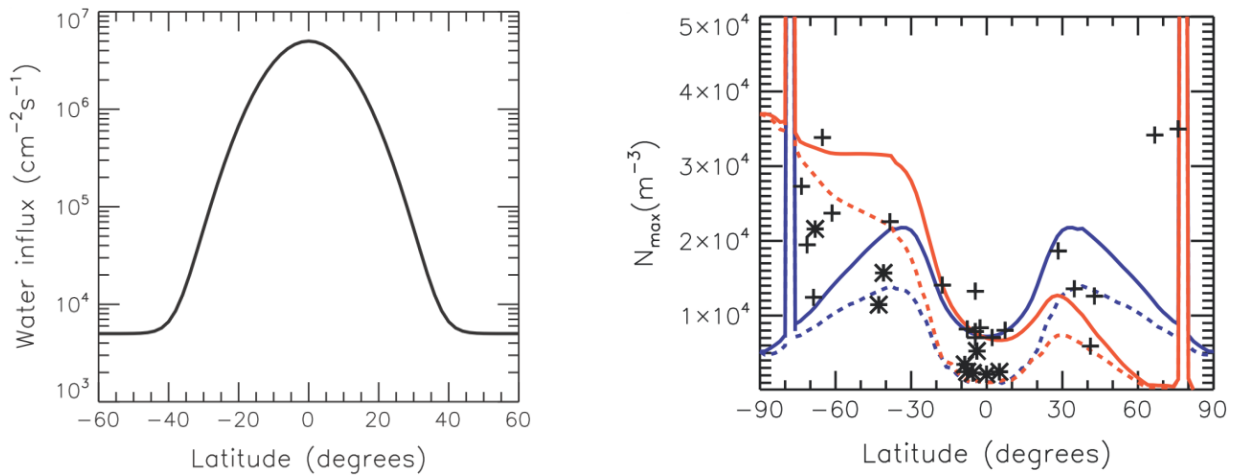

**Figure 8.13: (left)** Water influx imposed at the upper boundary of the vertical grid of STIM. A minimum base level of water influx is assumed poleward of ~40° in both hemispheres for numerical stability. **(right)** Latitudinal variation of peak electron densities in Saturn's ionosphere, as observed by Cassini, for dusk (plus symbols) and dawn (star symbols) conditions (*Nagy et al.*, 2006; *Kliore et al.*, 2009). Also shown are modeled peak electron densities for Saturn equinox (blue) and southern summer (red) at dusk (solid lines) and dawn (dashed lines). From *Müller-Wodarg et al.* (2012).

*Diurnal Variation of $N_{MAX}$*

The diurnal variation of the peak electron density, as inferred from Saturn Electrostatic Discharge (SED) observations, is between 1-2 orders of magnitude (**Figure 8.4**; *Kaiser et al.*, 1984; *Zarka*, 1985a; *Fischer et al.*, 2011b). Both Voyager and Cassini era SED measurements find an $N_{MAX}$ at noon of order $10^5$ cm$^{-3}$, with dawn and dusk values approximately $10^4$ cm$^{-3}$, in agreement with most radio occultations. Midnight electron densities derived from Voyager SEDs are ~$10^3$ cm$^{-3}$, while Cassini SEDs find a value nearer to $10^4$ cm$^{-3}$. This disagreement may arise from complications due to Saturn Kilometric Radiation (SKR) emission, it may be related to Saturn's ring shadows leading to locally depleted electron densities, or it may be attributed to something else entirely. At present, however, no model has come close to reproducing either the Voyager or the Cassini diurnal $N_{MAX}$ trend, regardless of the nighttime electron densities.

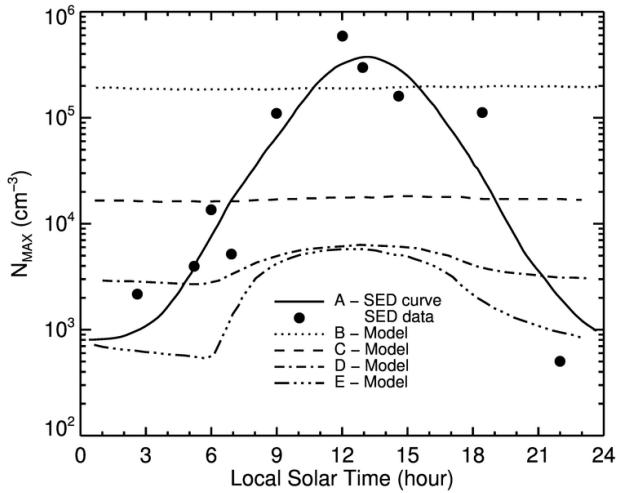

**Figure 8.14:** Modeled and SED- inferred diurnal variation of $N_{MAX}$. Curve A (solid line) shows the diurnal trend in peak electron density derived from SED measurements (filled circles). Curve B (dotted line) is the nominal model calculation, for which radiative recombination is the dominant loss of $H^+$. Curves C (dashed line), D (dot-dash line), and E (dash-triple-dot line) are model calculations that consider enhanced chemical losses due to water influx and vibrationally excited $H_2$. From *Majeed and McConnell* (1996).

*Majeed and McConnell* (1996) presented the first comprehensive model comparison with SED-derived ionospheric electron densities using a 1-D chemical diffusive model. They evaluated a wide range of combinations of vibrationally excited $H_2$ populations and water influxes in an attempt to reproduce the diurnal variation of $N_{MAX}$, as shown in **Figure 8.14.** *Majeed and McConnell* (1996) found that in the absence of these additional chemical losses the peak electron density was of order $10^5$ cm$^{-3}$ with little diurnal variation, in agreement with earlier ionospheric models. By enhancing the rate of plasma recombination, the modeled diurnal variation was also enhanced, though at the cost of a reduced $N_{MAX}$ that no longer accurately represented observed daytime electron densities. A subsequent study by *Moore et al.* (2012) focused on comparisons with $N_{MAX}$ derived from Cassini era SEDs. Rather than explore just a limited set of realistic vibrationally excited $H_2$ populations and water influxes, already shown not to work by *Majeed and McConnell* (1996), *Moore et al.* (2012) also considered enhanced ion production rates. Their goal was to answer the question, "What does it take to reproduce SED-derived diurnal variation?" **Figure 8.15** presents the result of 1-D model simulations that examined wide ranges of ion production and loss in order to reproduce the Cassini $N_{MAX}$ trend. The best fit model simulation required both drastic chemical loss and ion production in order to create and then destroy so many ions in only ~10 hours: the photoionization rate was increased by a factor of 60, the nominal population of vibrationally excited $H_2$ by a factor of 20, and the water influx by a factor of 540.

The primary difficulty in reproducing SED-derived diurnal electron density trends is the extremely rapid buildup of ionization in the morning hours implied by the measurements. For example, the net (i.e., production minus loss) electron production rate between dawn and noon from SED measurements is between ~9 cm$^{-3}$ s$^{-1}$ (Cassini era diurnal $N_{MAX}$ trend) and ~30-79 cm$^{-3}$ s$^{-1}$ (Voyager era diurnal $N_{MAX}$ trend), whereas the peak overhead production rate due to solar EUV is ~10 cm$^{-3}$ s$^{-1}$. Therefore, an explanation of SED observations may require some sort of extreme ionization process, such as due to a diurnal ionosphere-protonosphere exchange (e.g., *Connerney and Waite*, 1984). One alternative explanation is that SEDs may be sampling the narrow low-altitude layers frequently seen in radio occultation electron density profiles rather than the canonical "main" ionospheric peak. Such layers are consistent with the presence of gravity waves in Saturn's lower thermosphere (*Matcheva and Barrow*, 2012), and can lead to narrow regions of electron density enhancements without requiring any additional sources of ionization (*Barrow and Matcheva*, 2013). For this explanation to hold, however, these layers must also exhibit a strong diurnal variation, must correlate with solar flux (*Fischer et al.*, 2011b), and must be ever-present structures generated at a wide range of latitudes (*Moore et al.*, 2012). Finally, as the atmospheric storms that give rise to SEDs tend to occur only over a limited set of specific latitudes for currently unknown reasons (primarily 35°S; *Fischer et al.*, 2011b), any ionospheric explanation of SED-derived electron densities may also be local in nature.

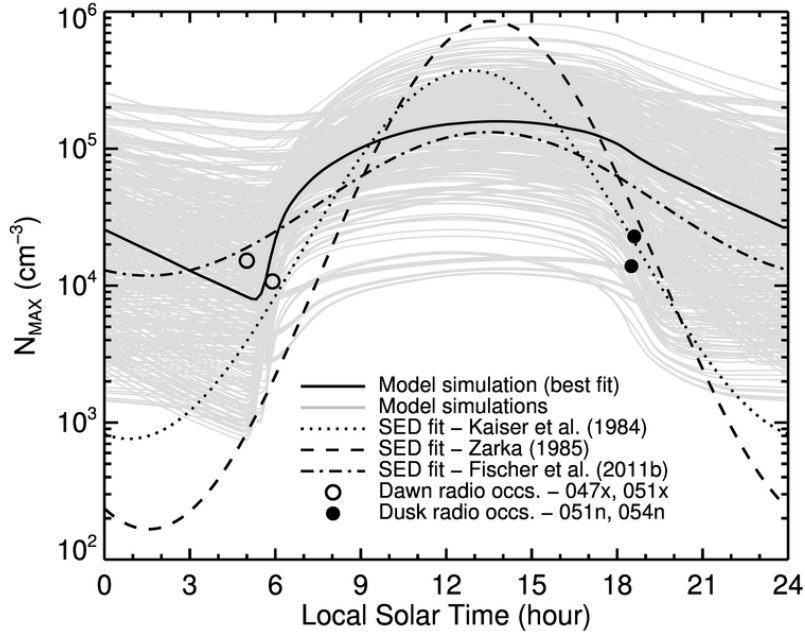

**Figure 8.15:** Model simulation (black solid line) that comes closest to reproducing the diurnal variation in $N_{MAX}$ derived from Cassini SEDs (dot-dash line). Also shown are the full range of model simulations (gray solid lines), the two Voyager era SED trends (dotted and dashed lines), the radio occultation measurements nearest to the Cassini era SED storm location at 35°S planetocentric latitude (filled circles for dusk, open circles for dawn). Both ion production and loss rates have been substantially increased over nominal model values. Adapted from *Moore et al.* (2012).

*Latitudinal $H_3^+$ Structure: Ionospheric Signatures of Ring Rain*

A new observational constraint was enabled recently following the first detection of $H_3^+$ emission at Saturn's non-auroral latitudes (*O'Donoghue et al.*, 2013), which revealed magnetically conjugate extrema in opposite hemispheres that also linked to structures in Saturn's rings. These low- and mid-latitude $H_3^+$ emission structures have therefore been interpreted as representing the ionospheric signatures of "ring rain", a process wherein charged water products from Saturn's rings are transported along magnetic field lines into its atmosphere (*Connerney*, 2013). Rather than being limited to electron density altitude profiles, like radio occultations, or to diurnal variations in peak electron density, like SED measurements, observations of ring rain signatures offer the possibility of deriving a snapshot of latitudinal variations in $H_3^+$ density and temperature near local noon. As one of the two major ions in the upper ionosphere, such a measurement could be combined with electron density trends from radio occultation measurements in order to estimate $H^+$ column densities, and thereby provide a significantly improved constraint on the loss processes that control $H^+$ densities – vibrationally excited $H_2$ populations and water influx.

Emission from $H_3^+$ ions depends exponentially upon their temperature. Typically, the ratios between various emission lines are used to derive the column-integrated temperature of the emitting gas, which further allows a calculation of the column density (e.g., *Melin et al.*, 2007). (Column-integrated temperature refers to the effective temperature derived for a column of gas that varies in both density and temperature along the column.) Unfortunately, due to a low signal-to-noise ratio, these parameters could not be derived from the only measured signatures of ring rain at present. Therefore, in order to make any direct model comparisons, a method of estimating the column densities is required.

In a recent study involving 1-D ionospheric calculations over a fixed 3-D background thermosphere, *Moore et al.* (2015) addressed this shortcoming by first estimating the neutral temperatures. They based latitudinal trends in exospheric temperature on solar and stellar EUV occultation results described in Chapter 9. As these temperatures represent conditions near the top of the atmosphere, above the altitude of peak $H_3^+$ ionization, an offset was then applied that accounted for the difference in neutral temperature between $H_3^+$ altitudes and the exobase as a function of latitude, as calculated by the 3-D STIM (*Müller-Wodarg et al.*, 2012). These temperature estimates were then combined with the measured intensities in order to derive $H_3^+$ column densities from the observations, shown in **Figure 8.16**. The structure in **Figure 8.16** is driven by the observed structure in $H_3^+$ emission, whereas the range of estimated column densities is dominated by temperature uncertainties.

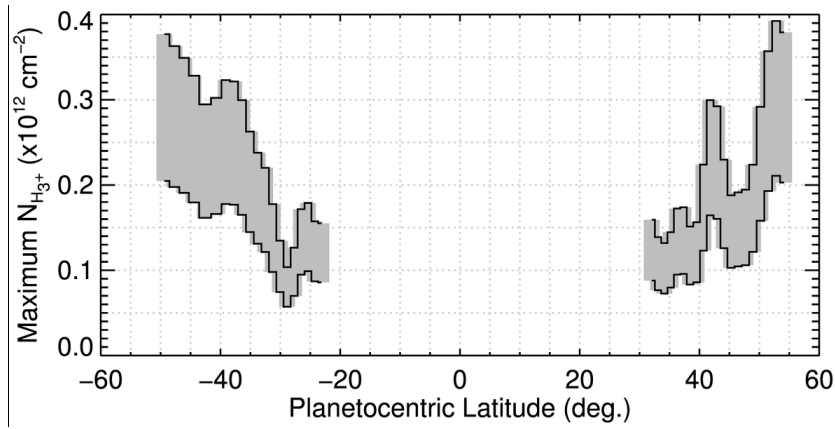

**Figure 8.16:** Maximum $H_3^+$ column densities estimated from Saturn observations of ring rain signatures (*O'Donoghue et al.*, 2013), based on exospheric temperature measurements (*Koskinen et al.*, 2013) and modeled temperature offsets (*Müller-Wodarg et al.*, 2012). The range of $H_3^+$ column densities is driven primarily by temperature uncertainties of ~80 K. Adapted from *Moore et al.* (2015).

While enhanced populations of vibrationally excited $H_2$ and water influxes have historically been considered as a means of reducing the modeled electron density (via destruction of the long-lived $H^+$ ions), there is an important secondary effect on $H_3^+$ densities. Specifically, as the dominant chemical loss for $H_3^+$ in outer planet ionospheres is due to dissociative recombination with electrons, the above loss processes also act as proxy "sources" of $H_3^+$ by reducing its dissociative recombination rate in correspondence with the reduced electron densities. This effect is augmented without limit by reactions between $H^+$ and vibrationally excited $H_2$, as they lead to additional production of $H_2^+$ (and therefore $H_3^+$), but it is reversed for water influxes above ~$10^7$ cm$^{-2}$ s$^{-1}$, as $H_3^+$ can also be lost due to charge exchange with water products (*Moore et al.*, 2015).

Contours of $H_3^+$ column density as a function of vibrationally excited $H_2$ population (designated as $k_{fac}$) and water influx are presented in **Figure 8.17**. A dashed contour indicates the estimated column density, and therefore represents the combinations of $k_{fac}$ and $\phi_{H_2O}$ that could reproduce the observation at that latitude, 35°S planetocentric. By performing a similar series of calculations for each of the 40 latitude elements of the observations, *Moore et al.* (2015) were able to derive a latitudinal variation in the maximum water influxes and maximum populations of vibrationally excited $H_2$. They estimated the globally averaged maximum ring-derived water influx to be $(1.6-12) \times 10^5$ cm$^{-2}$ s$^{-1}$, which represents a maximum total global influx of water from Saturn's rings to its atmosphere of $(1.0-6.8) \times 10^{26}$ s$^{-1}$. While there are no direct observational constraints for purely ring-derived water influxes, these upper limits compare favorably with the total oxygen influx of ~$10^5$ cm$^{-2}$ s$^{-1}$ estimated from ring atmosphere models (e.g., *Tseng et al.*, 2010). Furthermore, the globally averaged upper limits for water influx, when combined with the assumed neutral water influx profile of **Figure 8.13**, are (1.3-2.3)x$10^6$ cm$^{-2}$ s$^{-1}$, fairly close to the value of ~1.5x$10^6$ cm$^{-2}$ s$^{-1}$ derived by *Moses et al.* (2000) based on ISO observations (*Feuchtgruber et al.*, 1997). The wide ranges in these estimates stem primarily from uncertainties in $H_3^+$ column integrated temperatures, and therefore future observations of the ionospheric signatures of ring rain may be able to reduce those uncertainties by determining temperatures self-consistently.

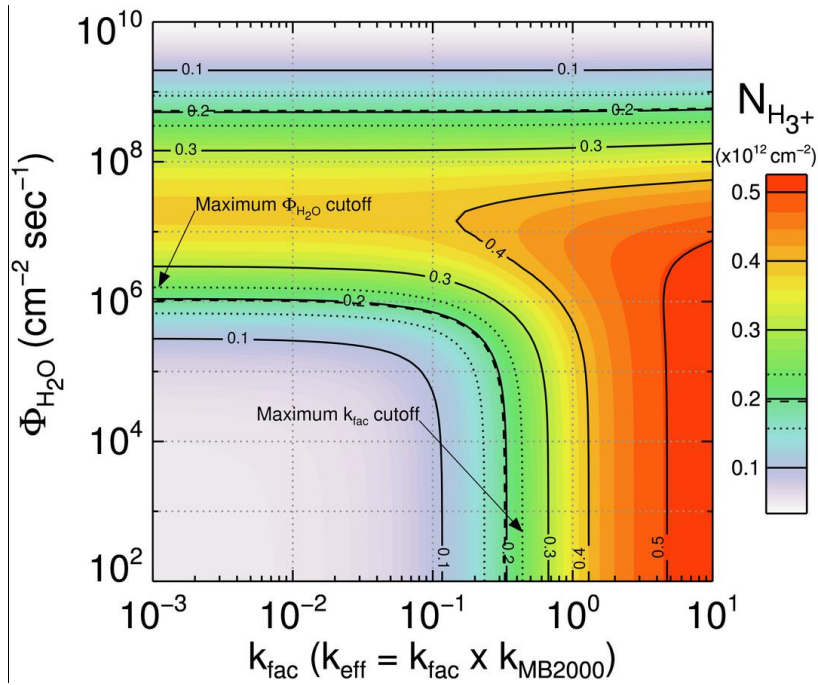

**Figure 8.17:** Contours of modeled $H_3^+$ column densities for a range of water influxes, $\phi_{H_2O}$, and populations of vibrationally excited $H_2$ (represented as $k_{fac}$). Calculation results are for 35°S planetocentric latitude at local solar noon. The dashed curves indicate the $H_3^+$ column density estimated from the observations (Figure 8.16), and therefore identify combinations of $\phi_{H_2O}$ and $k_{fax}$ that can reproduce the observation at that latitude. Dotted curves outline a range of $H_3^+$ column densities that result from accounting for a 3σ uncertainty in the observed emission intensity. From *Moore et al.* (2015).

## 8.4   Summary

Saturn ionospheric science has advanced significantly over the past decade. Prior to the arrival of Cassini, there were 6 electron density profiles from radio occultations – there are now a total of 65. The diurnal variation in peak electron density derived from Voyager era Saturn Electrostatic Discharge (SED) measurements was based on a few measurements over three SED episodes. In contrast, Cassini era diurnal variations of electron density were derived from 48 SED episodes by 2009, and there is a corresponding wider range of additional observational parameters to explore (e.g., Cassini distance from Saturn versus derived electron density). In addition, there is the promise of a new remote observational constraint for Saturn's mid- and low-latitudes: $H_3^+$ emissions and related ring rain implications. While the conclusions that can be drawn from the current measurements of ring rain signatures are not yet definitive, future observations should allow derivations of $H_3^+$ and $H^+$ column densities, column-integrated $H_3^+$ temperatures, and an in depth examination of the coupling between Saturn's rings and its atmosphere.

Ionospheric models are able to explain the gross ionospheric structure revealed by radio occultation measurements. Electron densities appear to be controlled primarily by a latitudinally varying water influx, reducing the need for enhanced populations of vibrationally excited molecular hydrogen. The observed dawn/dusk asymmetry is well explained by a mix of atomic ($H^+$) and molecular ($H_3^+$) ions forming Saturn's main ionospheric peak. Low-altitude layers are likely indicative of atmospheric gravity waves, which can lead to vertically varying plasma drifts.

The major remaining model-data discrepancy is in the diurnal variation of peak electron density. SED-derived electron densities feature large daytime values and strong diurnal variations. Models can reproduce one or the other of these constraints, but not both simultaneously, having thoroughly demonstrated that no combination of known chemistry can account for the SED-derived variations. Resolution of this discrepancy may lie in introducing additional dynamics to the modeling, such as a diurnal ionosphere-protonosphere exchange, or further investigation into the possibility that SEDs are sampling the narrow low-altitude layers of electron density, or something else entirely.

Future observations should help refine the modeling constraints in a number of important areas. Observations of ionospheric $H_3^+$ emissions can provide the first ion density and temperature measurements in Saturn's ionosphere. Additional insights into ion densities resulting from in situ measurements are expected during Cassini's end-of-mission proximal orbits in 2017. During this mission phase, called the Cassini Grand Finale, the spacecraft will sample Saturn's upper atmosphere directly, dipping below 2000 km altitude on 22 orbits, with periapse at low latitude near dusk. With luck, Cassini may also be able to measure near-Saturn SEDs or determine the properties of the particles on magnetic field lines connecting Saturn's rings to its atmosphere. Combined with the existing observational constraints, these two new datasets – ground-based observations and Cassini Grand Finale measurements – should help to refine the remaining model-data discrepancies and to pave the way for future fully coupled models of the Saturn system by providing much needed insight into ionospheric chemical losses due to water and to vibrationally excited $H_2$.

**Acknowledgements**

The authors wish to acknowledge the contribution of the International Space Sciences Institute (ISSI) in Bern, Switzerland, for hosting and funding the International Team on "Coordinated Numerical Modeling of the Global Jovian and Saturnian Systems", and the constructive discussions by colleagues attending the meetings. The manuscript benefitted significantly from thorough reviews by two anonymous reviewers. Support for this work at Boston University comes from NASA Grants NNX13AG21G, NNX13AG57G, and NNX14AG72G. M.G. was supported by the STFC of UK under grant ST/K001051/1 and ST/N000692/1. A.F.N. acknowledges support from the Cassini Mission as a member of the Radio Science Team.

**Arvydas J. Kliore**

This chapter is dedicated to the memory of Arvydas J. Kliore, a pioneer in the field of Radio Science. He was involved in spacecraft radio occultations from the very first NASA planetary mission, and has shaped our knowledge of Saturn's upper atmosphere in his role as Principle Investigator of the Cassini Radio Science Subsystem. Without him, this chapter would not exist.

28    *Moore, Galand, Kliore, Nagy & O'Donoghue*

32  *Moore, Galand, Kliore, Nagy & O'Donoghue*observations, *Icarus*, *221*(2), 508–516, doi:10.1016/j.icarus.2012.08.010.

Moore, L., J. O'Donoghue, I. Müller-Wodarg, M. Galand, and M. Mendillo (2015), Saturn ring rain: Model estimates of water influx into Saturn's atmosphere, *Icarus*, *245*, 355–366, doi:10.1016/j.icarus.2014.08.041.

Moore, L. E., M. Mendillo, I. C. F. Müller-Wodarg, and D. L. Murr (2004), Modeling of global variations and ring shadowing in Saturn's ionosphere, *Icarus*, *172*(2), 503–520, doi:10.1016/j.icarus.2004.07.007.

Moses, J., and S. Bass (2000), The effects of external material on the chemistry and structure of Saturn's ionosphere, *J. Geophys. Res.*, *105*(1999), 7013–7052.

Moses, J., B. Bézard, E. Lellouch, G. R. Gladstone, H. Feuchtgruber, and M. Allen (2000), Photochemistry of Saturn's Atmosphere II. Effects of an Influx of External Oxygen, *Icarus*, *145*(1), 166–202, doi:10.1006/icar.1999.6320.

Müller-Wodarg, I. C. F., M. Mendillo, R. Yelle, and A. Aylward (2006), A global circulation model of Saturn's thermosphere, *Icarus*, *180*(1), 147–160, doi:10.1016/j.icarus.2005.09.002.

Müller-Wodarg, I. C. F., L. Moore, M. Galand, S. Miller, and M. Mendillo (2012), Magnetosphere–atmosphere coupling at Saturn: 1 – Response of thermosphere and ionosphere to steady state polar forcing, *Icarus*, *221*(2), 481–494, doi:10.1016/j.icarus.2012.08.034.

Nagy, A. F., and T. E. Cravens (2002), Solar system ionospheres, in *Atmospheres in the Solar System: Comparative Aeronomy*, edited by M. Mendillo, A. Nagy, and J. H. Waite, pp. 39–54, American Geophysical Union, Washington, D.C.

Nagy, A. F., A. J. Kliore, E. Marouf, R. French, M. Flasar, N. J. Rappaport, A. Anabtawi, S. W. Asmar, D. Johnston, E. Barbinis, G. Goltz, and D. Fleischman (2006), First results from the ionospheric radio occultations of Saturn by the Cassini spacecraft, *J. Geophys. Res.*, *111*(A6), A06310, doi:10.1029/2005JA011519.

Nagy, A. F., A. J. Kliore, M. Mendillo, S. Miller, L. Moore, J. I. Moses, I. Müller-wodarg, and D. Shemansky (2009), Upper Atmosphere and Ionosphere of Saturn, in *Saturn from Cassini-Huygens*, edited by M. K. Dougherty, L. W. Esposito, and S. M. Krimigis, pp. 181–201, Springer Netherlands, Dordrecht.

Northrop, G., and J. R. Hill (1982), Stability of Negatively Charged Dust Graings in Saturn's Ring Plane, *J. Geophys. Res.*, *87*(A8), 6045–6051.

Northrop, T. G., and J. E. P. Connerney (1987), A Micrometeorite Erosion Model and the Age of Saturn's Rings, *Icarus*, *70*, 124–137.

Northrop, T. G., and J. R. Hill (1983), The Inner Edge of Saturn's B Ring, *J. Geophys. Res.*, *88*(A8), 6102–6108.

O'Donoghue, J., T. S. Stallard, H. Melin, G. H. Jones, S. W. H. Cowley, S. Miller, K. H. Baines, and J. S. D. Blake (2013), The domination of Saturn's low-latitude ionosphere by ring "rain," *Nature*, *496*(7444), 193–195, doi:10.1038/nature12049.

O'Donoghue, J., T. S. Stallard, H. Melin, S. W. H. Cowley, S. V. Badman, L. Moore, S. Miller, C. Tao, K. H. Baines, and J. S. D. Blake (2014), Conjugate observations of Saturn's northern and southern aurorae, *Icarus*, *229*, 214–220, doi:10.1016/j.icarus.2013.11.009.

Perry, J., Y. Kim, J. Fox, and H. Porter (1999), Chemistry of the Jovian Auroral Ionosphere, *J. Geophys. Res.*, *104*(E7), 16,541–16,565.

Porco, C. C., E. Baker, J. Barbara, K. Beurle, A. Brahic, J. A. Burns, S. Charnoz, N. Cooper, D. D. Dawson, A. D. Del Genio, T. Denk, L. Dones, U. Dyudina, M. W. Evans, B. Giese, K. Grazier, P. Helfenstein, A. P. Ingersoll, R. A. Jacobson, T. V Johnson, A. Mcewen, C. D. Murray, G. Neukum, W. M. Owen, J. Perry, T. Roatsch, J. Spitale, S. Squyres, P. Thomas, M. Tiscareno, E. Turtle, A. R. Vasavada, J. Veverka, R. Wagner, and R. West (2005), Cassini Imaging Science: Initial Results on Saturn's Atmosphere, *Science (80-. ).*, *307*(February), 1243–1247.

Porco, C. C., P. Helfenstein, P. C. Thomas, a P. Ingersoll, J. Wisdom, R. West, G. Neukum, T. Denk, R. Wagner, T. Roatsch, S. Kieffer, E. Turtle, A. McEwen, T. V Johnson, J. Rathbun, J. Veverka, D. Wilson, J. Perry, J. Spitale, A. Brahic, J. a Burns, a D. Delgenio, L. Dones, C. D. Murray, and S. Squyres (2006), Cassini observes the active south pole of Enceladus, *Science (80-. ).*, *311*(5766), 1393–401, doi:10.1126/science.1123013.

Prangé, R., T. Fouchet, R. Courtin, J. E. P. Connerney, and J. C. McConnell (2006), Latitudinal variation of Saturn photochemistry deduced from spatially-resolved ultraviolet spectra, *Icarus*, *180*(2), 379–392, doi:10.1016/j.icarus.2005.11.005.

Schlesier, A. C., and M. J. Buonsanto (1999), The Millstone Hill ionospheric model and its application to the May 26-27, 1990, ionospheric storm, *J. Geophys. Res.*, *104*(A10), 22,453–22,468.

Schunk, R. W., and A. F. Nagy (2009), *Ionospheres: Physics, Plasma Physics, and Chemistry*, 2nd ed., Cambridge University Press, Cambridge, UK.